\documentclass[aoas,MSNbibl,nameyear,seceqn,rotating,dvips]{arximspdf}
\usepackage{graphicx}

\doi{10.1214/10-AOAS441}
\volume{5}
\issue{2B}
\pubyear{2011}
\firstpage{1293}
\lastpage{1327}

\makeatletter
\renewcommand{\cite}{\citet}
\newcommand{\eqref}[1]{(\ref{#1})}
\renewcommand{\epsilon}{\varepsilon}

\newcommand{\E}{\mathrm{E}}
\newcommand{\ODF}{\operatorname{ODF}}
\newcommand{\be}{\mathbf{e}}

\newcommand{\bq}{\mathbf{q}}

\newcommand{\q}{\mathbf{q}}
\newcommand{\tbq}{\tilde{\bq}}

\newcommand{\bu}{\bolds{\upsilon}}
\newcommand{\uu}{\mathbf{u}}
\newcommand{\x}{\mathbf{x}}

\newcommand{\cA}{\mathcal{A}}

\newcommand{\cG}{\mathcal{G}}

\newcommand{\cQ}{\mathcal{Q}}
\newcommand{\R}{\mathcal{R}}

\newcommand{\bsL}{\bolds{\Lambda}}
\newcommand{\bsU}{\bolds{\Upsilon}}
\newcommand{\bld}[1]{\mathbf{#1}}
\newcommand{\wh}[1]{\widehat{#1}}
\newcommand{\wt}[1]{\widetilde{#1}}
\newcommand{\ol}[1]{\overline{#1}}
\makeatother

\begin{document}
\begin{frontmatter}

\title{Nonparametric tests of structure for high angular resolution diffusion imaging in $Q$-space\protect\thanksref[1]{T1}}
\runtitle{Nonparametric Testing for HARDI}
\thankstext{T1}{Supported by EPSRC and GlaxoSmithKline via EP/E031536/1.}
\begin{aug}
\author[A]{\fnms{Sofia C.} \snm{Olhede}\ead[label=e1]{s.olhede@ucl.ac.uk}}
\and
\author[B]{\fnms{Brandon} \snm{Whitcher}\corref{}\ead[label=e2]{brandon.j.whitcher@gsk.com}}

\runauthor{S. C. Olhede and B. Whitcher}
\affiliation{University College London and GlaxoSmithKline}
\address[A]{Departments of Computer Science\\
\quad  and Statistical Science\\
University College London\\
Gower Street\\
London WC1 E6BT\\
United Kingdom\\
\printead{e1}} 
\address[B]{GlaxoSmithKline Clinical\\
\quad Imaging Centre\\
Hammersmith Hospital\\
Imperial College London\\
Du Cane Road\\
London W12 0HS\\
United Kingdom\\
\printead{e2}}
\end{aug}

\received{\smonth{6} \syear{2010}}
\revised{\smonth{10} \syear{2010}}

\begin{abstract}
High angular resolution diffusion imaging data is the observed
characteristic function for the local diffusion of water molecules in
tissue. This data is used to infer structural information in brain
imaging.  Nonparametric scalar measures are proposed to summarize
such data, and to locally characterize spatial features of the
diffusion probability density function (PDF), relying on the geometry
of the characteristic function.  Summary statistics are defined so
that their distributions are, to first-order, both independent of
nuisance parameters and also analytically tractable.  The dominant
direction of the diffusion at a~spatial location (voxel) is
determined, and a~new set of axes are introduced in Fourier space.
Variation quantified in these axes determines the local spatial
properties of the diffusion density.  Nonparametric hypothesis tests
for determining whether the diffusion is unimodal, isotropic or
multi-modal are proposed.  More subtle characteristics of white-matter
microstructure, such as the degree of anisotropy of the PDF and
symmetry compared with a variety of asymmetric PDF alternatives, may
be ascertained directly in the Fourier domain without parametric
assumptions on the form of the diffusion~PDF.  We simulate a set of
diffusion processes and characterize their local properties using the
newly introduced summaries.  We show how complex white-matter
structures across multiple voxels exhibit clear ellipsoidal and
asymmetric structure in simulation, and assess the performance of the
statistics in clinically-acquired magnetic resonance imaging data.
\end{abstract}


\end{frontmatter}

\section{Introduction}
\label{intro}

\setcounter{footnote}{1}
Many applications in brain imaging are based on calculating local
statistics that are later combined to infer global properties of
spatial links or functional connections.  In this paper we focus on the
local analysis of high angular resolution diffusion imaging (HARDI)
data, a~special type of magnetic resonance imaging (MRI). HARDI
observations correspond to the local (in a single voxel\footnote{A
voxel is a three-dimensional ``volume element'' of
  data, just as a pixel is a two-dimensional ``area element'' of
  data.}) measurement of the local molecular diffusion of water at a
number of different orientations over a spherical shell of fixed radius
[\cite{Callaghan}].  Measurements from an MRI scanner are taken
directly in the Fourier domain and translated into the spatial domain
via the inverse Fourier transform.

A HARDI acquisition scheme permits the characterization of directional
spatial properties of the diffusion probability density function (PDF).
The local structure of white-matter brain tissue may be inferred from
such measurements [\cite{basser1994}; \cite{basrelationships}]. Once local
statistics have been formed, it is of interest to combine information
across voxels (spatial locations), for example, to connect local
directions of estimated diffusion~PDFs to recognize major nerve fiber
tracts, to infer local fiber structure from the estimated diffusions
[\cite{morzijfibertracking}], and/or to use other locally-defined
statistical summaries in inferential procedures
[\cite{jenetalkurtosis}].

Different orientational sampling designs can be used at each voxel and,
if a~simple parametric model is used for the PDF, then rather sparse
sampling will be sufficient to recover the parameters of the model.
Traditional analysis of HARDI measurements is based on modeling the
diffusion~PDF parametrically as a (zero-mean) Gaussian, and estimating
a diffusion tensor (the covariance matrix of the Gaussian PDF), a
procedure which corresponds to diffusion tensor imaging (DTI).  Such
methods have drawbacks, namely, of not describing more complex
white-matter structures well, and their usage trades a small variance
for potentially large bias.  While the diffusion tensor model has both
theoretical justification---and has been extremely popular---it
prohibits one from describing more complicated white-matter
microstructure, such as crossing, kissing and forking fibers
[\cite{morzijfibertracking}].

It is believed that intravoxel orientational heterogeneity affects as
many as one third of all imaged white-matter voxels
[\cite{Behrens2007}], and so addressing such structure is important.
With more time-intensive sampling schemes (such as HARDI
[\cite{tucetalhigh}] or diffusion spectrum imaging), the possibility
of more complicated estimators may be used, for example, multi-tensor
modeling [\cite{Alexander2005}], nonparametric alternatives such as
persistent angular structure MRI [\cite{Jansons}], $Q$-ball imaging
[\cite{Tuch}], the diffusion orientation transform
[\cite{Ozarslan2006}] and spherical deconvolution [\cite{Tournier}].
While using a nonparametric approach removes bias, usage of such
nonparametric methods is challenging because the diffusion process is
measured in the Fourier domain ($q$-space\footnote{$Q$-space is the
  Fourier domain representation of the local diffusion and is the
  space where measurements are made in MRI.  The global image Fourier
  representation is usually inverted to a spatial representation, but
  the local Fourier transform is not inverted as part of the
  acquisition, leaving the spatial domain observations associated with
  a measurement of local diffusion in a Fourier domain orientation.}),
and the characteristic function has been considerably undersampled to
accommodate realistic scanning times in practice.  This challenges the
stable inversion of information, the local characteristic function, to
local spatial structure.

\begin{figure}

\includegraphics{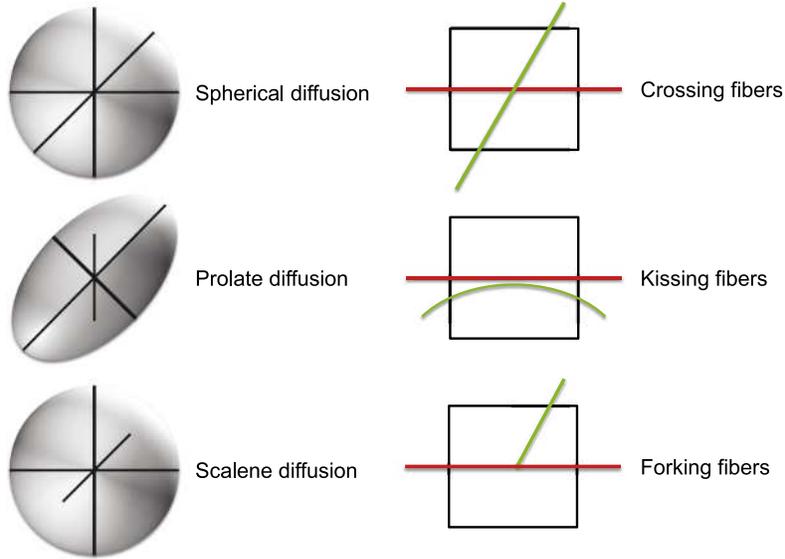}

  \caption{Simplified diagrams for typical Gaussian diffusion models
    \textup{(first column)} and fiber configurations in a voxel of white matter
    in the brain \textup{(second column)}.  Spherical diffusion is found when
    no fibers are present in a voxel of brain tissue (e.g.,
    cerebral-spinal fluid) and all eigenvalues are equal
    $(\lambda_1=\lambda_2=\lambda_3)$.  Prolate diffusion is when a
    single fiber bundle is present in the voxel
    $(\lambda_1\gg\lambda_2=\lambda_3)$.  Scalene diffusion is when
    two fiber bundles of similar mass cross in perpendicular
    directions $(\lambda_1\approx\lambda_2\gg\lambda_3)$.  The concept
    of ``crossing fibers'' involves two fiber bundles that do not
    necessarily intersect at right angles in the same voxel.  The
    concept of ``kissing fibers'' involves two fiber bundles that
    occupy the same voxel, but do not intersect.  The concept of
    ``forking fibers'' involves a single fiber going in the voxel and
    two fiber bundles leaving the voxel.  A ``fanning fiber'' (not
    shown) is similar to a forking fiber, but instead of a single
    direction the fiber produces multiple diverging fibers on one side
    of the voxel.} \label{fig:introduction}
    \vspace*{6pt}
\end{figure}

This paper develops a statistical framework, using nonparametric
methods, for characterizing HARDI data directly in $q$-space
[\cite{tucetalhigh}] without local inversion.  This avoids
calculating nonlinear transformations of the data, whose usage usually
leads to intractability of the distributions of statistical summaries.
The approximate distributions of the proposed estimators in this paper
are derived and are defined so that, to first order, they are free of
any nuisance parameters.  The proposed statistics are a first step
toward the automated detection of subtle characteristics of
white-matter microstructure, that is, scalene diffusions
(Figure~\ref{fig:introduction}) or asymmetry in decay in a fixed axis.
Both properties, scalene diffusion and asymmetry, have been found in a
forking fiber structure (Figure~\ref{fig:introduction}), and may be
important summaries to feed into fiber-tracking algorithms
[\cite{morzijfibertracking}].  The derived methods also serve as a
warning when interpreting multi-tensor models in clinically-feasible
acquisition schemes, as similar characteristics can be obtained from
more complex single peaked structures.

Global features like bi- or multi-modality of the diffusion~PDF are
described reasonably well by many methods over a range of
signal-to-noise ratios (SNRs), with the small caveat that the various
implicit assumptions inherent to any of the given methods must be
satisfied.  Parametric models introduce bias when they are not
appropriate, whereas using a nonparametric method increases the
variance in the estimation.  Using a moderate number of directions in
the HARDI sampling scheme restricts the possibility of determining
smaller scale structure of the diffusion~PDF.  Strong parametric
assumptions increase the power of any proposed statistic to detect
multiple diffusion directions, with the consequence that any deviation
from the prescribed structure in the parametric model may be used to
reject null hypotheses such as unimodality.

In the method proposed here to determine the properties of the
diffusion PDF, prolate diffusion PDFs are separated from isotropic (or
spherical) PDFs using a~test based on a comparison of relative
magnitudes in $q$-space; see Figure~\ref{fig:introduction} for
illustrations of prolate and spherical diffusion models. Subsequently,
multi-modal distributions are then differentiated from the isotropic
and unidirectional.  The unidirectional diffusion is associated with a
great circle in $q$-space [\cite{Tuch}], and we call this the
\textit{dominant great circle}.  The strongest direction defines an
important spatial summary of the diffusion PDF, and specifies the major
axis of the diffusion in $q$-space (Figure~\ref{fig:inrha}g).  The
perpendicular to the major direction in space defines a set of points
lying on a great circle in $q$-space, which exactly corresponds to the
dominant great circle.

If a given voxel has been diagnosed as unidirectional (or if there is
a~dominant great circle in $q$-space), then we seek to characterize its
main unidirectional structure in more detail.  A scalar measure
resembling the popular fractional anisotropy\footnote{The fractional
  anisotropy (FA) is a measure of uniformity of the eigenvalues of
  a~Gaussian covariance matrix [\cite{basFA}].} is defined as the
\textit{anisotropy statistic}, by comparing the magnitude of the
$q$-space diffusion on the dominant great circle with its two
perpendicular point(s).  This measure determines the degree of
anisotropy of the diffusion~PDF.  Further investigation of
unidirectional voxels causes us to focus on quantifying the uniformity
of decay in the minor axes of the diffusion~PDF, or the perpendicular
to the dominant great circle, to describe further detailed structure of
the characteristic function.

Ellipsoidal diffusions are an important class of diffusions and the
scalene structure of the diffusion~PDF is particularly important when
combining voxel-wise information [\cite{Seunarine}].  The
aforementioned work showed that the scalene structure of the peak is
related to the peak anisotropy in space and important for treating
bending and fanning fibers (Figure~\ref{fig:introduction}).
For diffusions with ellipsoidal decay, their minor axes are well
defined by this (scalene) decay structure, while for nonellipsoidal
diffusions the minor axes correspond to a set of axes in the plane of
the dominant great circle, parameterizing locations on the dominant
great circle.  We examine the scalene structure of the diffusion~PDF,
which is quantified by the difference in decay in the two spatial minor
axes, defined as such also for nonellipsoid diffusions.  This
corresponds to examining the variability of the diffusion on the great
circle perpendicular to the vector associated with the major direction
of the diffusion.  For a Gaussian diffusion model this is given by the
two minor eigenvalues of the eigen-decomposition of the diffusion
tensor.  A statistical test for uniformity on the great circle is
developed that can be related to the spatial decay of the diffusion~PDF
in the minor axes.  Another feature of interest in the PDF is asymmetry
in the decay in a fixed direction perpendicular to the dominant great
circle.  This heuristic may be visualized in space as a diffusion~PDF
that appears ellipsoidal but the peak is in one of the foci rather than
the center of the ellipse. We introduce a~test statistic for asymmetry
based on this understanding.  To motivate our interest in asymmetry and
ellipsoidality, we simulate forking and crossing structures, and show
how both asymmetry and ellipsoidality follow as precursors to forking
structure, and such information could be used to improve the tracking
of fibers.

The methodology presented here improves our understanding of the
diffusion~PDF by not relying on parametric assumptions when analyzing
the measurements, yet still relating $q$-space structure directly to
spatial properties.  Nonparametric statistical summaries are defined
directly in $q$-space to increase the power of the proposed hypothesis
tests and theoretical critical values for the statistics are provided.
Understanding the inherent limitations of HARDI measurements can be
obtained directly from our discussion of simulated diffusions, thus
increasing the understanding of parametric assumptions that are
necessary to derive more complicated structures from the diffusion~PDF.

\section{Statistical models for HARDI data}

\subsection{Observational model}

We denote the sampling of the observations by the set
$\cQ_0=\{\tbq_i\}_{i=1}^n$.  At each
$\tbq_i=(\tilde{q}_{i1},\tilde{q}_{i2},\tilde{q}_{i3})$ on the unit
sphere
$\|\tbq\|=(\tilde{q}_{1}^2+\tilde{q}_{2}^2+\tilde{q}_{3}^2)^{1/2}=1$ we
obtain an observed measurement $\wt{A}(\tbq_i)\ge0$, corresponding to
the magnitude of a complex-valued observation (proportional to the
noisy characteristic function of local diffusion\footnote{Note that
  this is different from the empirical characteristic function.}).
Furthermore, we take~$n_0$ observations at $\q=\bld{0}$, denoted by
$\wt{A}_k(\bld{0})$ for $k=1,\dots,n_0$.  We distinguish here between
the measured apparent diffusion at $\tbq_i$, namely, $\wt{A}(\tbq_i)$,
and the theoretical diffusion value, $\cA(\tbq_i)$. Note that the
expected value of $\wt{A}(\tbq_i)$ is \textit{not} equivalent to
$\cA(\tbq_i)$, for two reasons. First because the observations are
magnitudes, with the noise contributing in the expectation, and second
we need to re-normalize the observed diffusion to have unit volume, as
noted by \cite{Alexander2005}.  As the PDF is a density, it has to
satisfy the normalization of
%
\begin{equation}
  \int\!\!\! \int\!\!\! \int a(\x)\,d^3\x = 1\quad\Rightarrow\quad\cA(\bld{0}) = 1,
\end{equation}
where $a(\x)$ is the diffusion probability density function (PDF), or
the inverse Fourier Transform of $\cA(\q)$.  We apply a biased
estimator of a simple average to estimate the inverse of the
normalizing constant by
$\ol{A}(\bld{0})=n_0^{-1}\sum_{k=1}^{n_0}\wt{A}_k(\bld{0})$.  We
re-normalize the observed diffusion such that
$A(\tbq_i)=\wt{A}(\tbq_i)/\ol{A}(0)$.  The diffusion value $A(\tbq_i)$
has (approximately) a Rician distribution with parameters $\cA(\tbq_i)$
and $\sigma^2$ [\cite{Gudbjartsson}].  As the SNR will be large at
$\q=\bld{0}$, the noise floor of the Rician distribution will have
limited impact in the estimation of the normalization constant. While
the diffusion PDF $a(\x)$ is not Gaussian, the Rician distribution
under reasonable SNR is well approximated by the Gaussian, and sums of
Rician variables will be very similar to a Gaussian.  In subsequent
sections we shall calculate statistical estimators from normalized
measurements $\{A(\tbq_i)\}_i^n$ and look at maxima of these
statistics, which may be represented (approximately) by the maxima of
suitably-scaled Gaussian random variables.  If we are in the regime of
low SNR, then these test statistics will be approximated by a~mixture
of Gaussian and Chi random variables whose tail-behavior is not
substantially heavier than Gaussian random variables, but whose mean is
not consistent with our results.  An~assumption for the method to work
is therefore a reasonable level of the SNR, as is further discussed in
Section \ref{sec:simulation}.

The normalized diffusion measurements $\cA(\tbq_i)$ should exhibit
symmetry as the diffusion PDF is real-valued, symmetric and indeed
positive, that is, $\cA(-\tbq_i)=\cA(\tbq_i)$ [\cite{Wedeen05}].  To
fully exploit the Hermitian symmetry, we shall reflect the observations
to the augmented set $\cQ=\{\q\dvtx\q\in\cQ_0\}\cup\{\q\dvtx-\q\in\cQ_0\}$, and
set $A(-\tbq_i)=A(\tbq_i)$ [\cite{Jansons}].

We assume that a nonparametric estimator of the diffusion in $q$-space
is constructed.  For our purposes we have chosen to use a
variable-bandwidth estimator
[Olhede and Whitcher (\citeyear{OlhedeWhitcher}, \citeyear{OlhedeWhitcherISBI})], but the methodology
outlined here is applicable to other linear estimators (e.g., radial
basis functions and/or spherical harmonics) with some straightforward
alteration of the statistical properties (specifically, second-order
structure) of the estimators.

\subsection{Great circles in $q$-space}

Spatial properties of the diffusion~PDF may be described directly in
$q$-space.  The advantage of such an operation is that we avoid the
need to invert the PDF to the spatial domain for analysis, allowing us
to employ a broad range of modeling approaches.  A basic building block
of our analysis is an \textit{ellipsoid density}.  We refer to a
density $a_\mathrm{E}(\x)$ as an ellipsoid density if its FT takes the form
\begin{equation}\label{ellipsoid}
  \cA_\mathrm{E}(\q; \bsL,\bsU) =
  B\Biggl(\sqrt{\sum_{j=1}^3\lambda_j|\bu_j^T\q|^2
  }\Biggr),
\end{equation}
where $\lambda_j\ge{0}$ for $j=1,2,3$, $\{\bu_j\}$ constitutes a basis
for ${\mathbb{R}}^3$ and $B(\cdot)$ is a~monotonically decreasing
function.  For example, it is common to use the Gaussian characteristic
function $B(q)=e^{-2(\pi q)^2}$.  We collect the eigenvalues in the
matrix
$\bsL=\operatorname{diag}(\lambda_1,\lambda_2,\lambda_3)$, and
define
%
\begin{equation}
  \bsU^T = \left[\matrix{
         \upsilon_{11} & \upsilon_{12} & \upsilon_{13}\cr
      \upsilon_{21} & \upsilon_{22} & \upsilon_{23}\cr
      \upsilon_{31} & \upsilon_{32} & \upsilon_{33},
   }\right]
\end{equation}
to model the axis of any orientational structure.  Ellipsoid densities
are natural building blocks, just like the special case of the DTI
model, but do not (for example) include multi-modal densities.  If the
$q$-space density takes this form, then the spatial PDF is given by
inverting the FT
%
\begin{equation}
  a_\mathrm{E}(\x;\bsL,\bsU) = \int\!\!\!\int\!\!\!\int_{\mathbb{R}^3}
  \cA_\mathrm{E}(\q; \bsL,\bsU)e^{i2\pi \q^T\x}\,d^3 \q
\end{equation}
[\cite{Callaghan}].  We note for $\x\in\mathbb{R}^3$, with $x=\|\x\|$
and $q=\|\q\|$, that $a_\mathrm{E}(\x;\bsL,\bsU)$ takes the form
%
\begin{equation}\label{ellipsoid2}
  a_\mathrm{E}(\x;\bsL,\bsU) = |\bsL|^{1/2}
  b(\|\bsL^{-1/2}\bsU\x\|),
\end{equation}
where
\begin{eqnarray}
  b(x) &=& \int_{-\infty}^{\infty} \int_{-\infty}^{\infty}
  \int_{-\infty}^{\infty} B(q) e^{i2\pi\q^T\x}\,d^3\q\\
  &=& \frac{1}{2\pi^2 x}\int_0^{\infty}
  B\biggl(\frac{q'}{2\pi}\biggr)\sin(xq')q'\,dq'\\
  &=& \frac{2}{x}\int_0^{\infty}B(q)\sin(2\pi xq) q\,dq,
\end{eqnarray}
which follows from \cite{grad}, page~1112.  The meaning of ``ellipsoid
density'' becomes clear from this expression, since whenever
$\|\bsL^{-1/2}\bsU\x\|=R$, where $R\ge 0$ is a constant, the function
$a_\mathrm{E}(\x)$ takes the same value in space.  As long as all the
eigenvalues are positive, $ a_\mathrm{E}(\cdot)$ will map out ellipsoidal
contours of equal function value in space.  The Gaussian DTI model fits
into this class of densities with $b(x)=(2\pi)^{-3/2}e^{-x^2/2}$ as
well as, for example, the Mat\'{e}rn family with the spatial variable
exchanged with the spatial-frequency variable [\cite{Matern}].  The
model proposed by \cite{Kaden2007} is also related to such densities.

\begin{figure}

\includegraphics{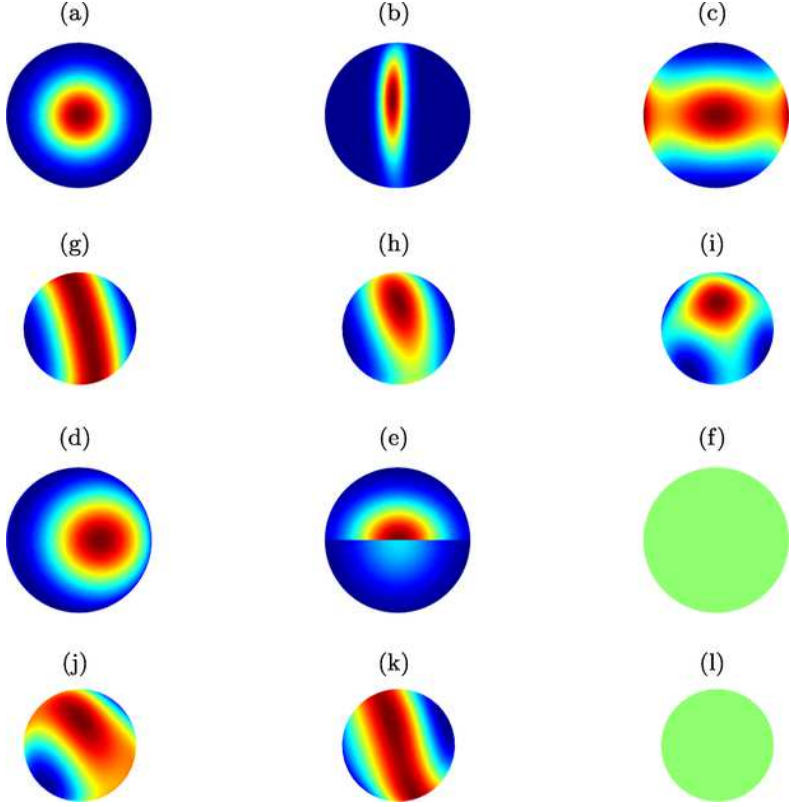}

  \caption{Diffusion processes displayed in both spatial and
    frequency ($q$-space) domains, with coloring representing density
    on the sphere.  Ellipsoid diffusions are represented by their
    covariance matrix eigenvalues $\{\lambda_i\}_{i=1}^3$ which govern
    a symmetric spatial decay.  \textup{(a,~g)} Prolate (ellipsoid)
    diffusion process $(\lambda_1\gg\lambda_2=\lambda_3)$.  A prolate
    diffusion process is dominated by a single direction, represented
    by \textup{(a)} a single peak in the diffusion~PDF and \textup{(g)} a great circle
    perpendicular to the diffusion direction in $q$-space.
    \textup{(b, h)} Scalene (ellipsoid) diffusion process
    $(\lambda_1\approx\lambda_2\gg\lambda_3)$.  A scalene diffusion
    process has two competing directions, which makes the minor axes
    unequally matched in both spaces.  \textup{(c, i)} A mixture of
    prolate (ellipsoid) diffusions.  This cannot be represented by a
    single unimodal diffusion~PDF but must be represented by two
    directions. \textup{(d, j)} and \textup{(e, k)} These are both
    (nonellipsoid appearing) diffusion~PDFs with asymmetric
    structure, suitable to model precursors to branching or forking
    (see text).  Neither of these diffusion~PDFs can be thought of as
    ellipsoid.  \textup{(f, l)} Isotropic diffusion with no directional
    structure in space or $q$-space.}
  \label{fig:inrha}
\end{figure}

Figure~\ref{fig:inrha} provides examples of diffusion processes
displayed in both the spatial and frequency ($q$-space) domains.  The
spatial domain corresponds to the diffusion~PDF, whereas its Fourier
transform corresponds to the $q$-space representation.  Common
processes, such as prolate and scalene diffusion, are given as well as
more exotic examples, such as a mixture of prolate diffusion processes
and a~process that cannot be represented using a Gaussian diffusion
model.  The values of $\bsU$ specify the orientation of the
diffusion~PDF, while $\bsL$ gives its qualitative appearance when
coupled with $B(\cdot)$.  Looking directly at Figure~\ref{fig:inrha},
it may be difficult for one to appreciate the local structure near the
peak, which motivates us to develop a new class of statistics to
characterize the diffusion~PDF.

\subsection{The orientation distribution function}

An important tool in understanding HARDI data is the orientational
distribution function (ODF).  The ODF quantifies the directional
structure of the diffusion~PDF in space.  A~popular object of study, it
corresponds to several different functions in the literature.
\cite{Tuch} and \cite{Hess}; \cite{Descoteaux} define the ODF to be
%
\begin{equation}
  \ODF_\mathrm{T}(\theta,\phi) = \frac{1}{Z} \int_0^{\infty} a(r\uu)\,dr,
\end{equation}
where $\x=r\uu$, $\|\uu\|=1$ and $Z$ is a normalizing constant. Because
this is not a true marginalization of a PDF (the increment needs a
weighting by~$r^2$), and weights lower scales heavily, the diffuse
directional structure of the large-scale structure smooths the
marginal~PDF of orientations, giving it a~``blunted'' appearance.  A~nonlinear transformation is necessary for the ODF to have a more peaked
and clear directional structure.  \cite{Wedeen05} define the ODF as
the truly marginalized PDF over all spatial radii
%
\begin{equation}
  \ODF_\mathrm{W}(\theta,\phi) = \int_0^{\infty} r^2 a(r\uu)\,dr.
\end{equation}
An alternative version may be found in \cite{Jansons}, where the
orientational structure associated with a single radius is fitted to
the observed data, that is, the persistent angular structure (PAS-MRI)
algorithm.  It is useful to note that the observed data are not
associated purely with a single radius, and for this to be a
mathematically correct procedure the observed HARDI measurements should
be convolved with a suitable kernel prior to estimation. Despite this
fact, the PAS-MRI method usually produces good results in practice.
All three of these orientational summaries are measuring different
properties of the directional structure of the data, and only
$\ODF_\mathrm{W}(\cdot,\cdot)$ is a true marginal PDF.

Another directional representation of diffusion data corresponds to the
spherical convolution model [\cite{Tournier}].  In this model,
$q$-space observations are modeled as convolved fiber ODFs, and fiber
populations are estimated using deconvolution methods.  The magnitudes
are not comparable with previously-defined estimators of ODFs.
Extensions to these methods have also been proposed: by modeling the
ODF as a mixture of Bingham distributions [\cite{Kaden2007}], and by
regularizing the deconvolution problem by applying constrained
optimization methods [\cite{Jian2007}].  The solution in
\cite{Kaden2007} is parametric and the theoretical assumptions
necessary to apply the regularized methods are, in general, violated
[\cite{Jian2007}].

The ellipsoid diffusion model \eqref{ellipsoid2} may be extended into a
larger class of arbitrarily peaked and deformed diffusion~PDFs by
taking
%
\begin{equation}
  \bsL(\x) =
  \operatorname{diag}(\lambda_{11}(\x),\lambda_{22}(\x),\lambda_{33}(\x)),
  \qquad \lambda_{jj}(\x) \ge 0  \forall \x,
\end{equation}
with $C$ a normalizing constant, to produce the diffusion~PDF
\begin{eqnarray}
  a_{\mathrm{DE}}(\x) &=& C \sqrt{|\bsL(\bsU\x)|}
  b(\|\bsL(\bsU\x)^{-1/2}
  \bsU\x\|),\\
  a_{\mathrm{DE}}(\bsU^T\x) &=& C \sqrt{|\bsL(\x)|}
  b(\|\bsL(\x)^{-1/2}\x\|).
\end{eqnarray}
Because $\bsL(\x)$ is a diagonal matrix, $a_{\mathrm{DE}}(\bsU^T\x)$ exhibits
the axes $(1,0,0)$, $(0,1,0)$ and $(0,0,1)$.  Applying a Fourier
transform directly, with a change of variables, we note that the
Fourier transform is mixed over the strengths in~$\bsL(\x)$, but
exhibits the same orientational axes if the ordering in magnitude of
the eigenvalues does not switch over $\x$.  We have the model of
%
\begin{equation}\label{deformedell}
  \cA_{\mathrm{DE}}(\q) = C \int\!\!\!\int\!\!\!\int_{\mathbb{R}^3} |\bsL(\x)|^{1/2}
  b(\|\bsL(\x)^{-1/2}\x\|)  e^{-i2\pi(\bsU\q)^T\x}\,d^3\x.
\end{equation}
This function can take the appearance of a deformed ellipsoid in space,
and may then exhibit a different pattern of decay to the left and right
of the dominant great circle in $q$-space.  For the regular ellipsoid
distribution $a_\mathrm{E}(\x)$ if one eigenvalue is larger than the two others
(say, $\lambda_1>\lambda_2\ge\lambda_3$), then the ellipsoid density
[or equally in the case of the deformed density if
$\inf_{\x}\lambda_1(\x)>\sup_{\x}\lambda_2(\x)$] will observe a
maximum at the values 
\begin{equation}\label{qbeta}
  \hspace*{15pt}\q(\beta) = \cases{
    \beta \bu_2+\sqrt{1-\beta^2}\bu_3, &\quad if
    $\beta \in [-1,1]$,\vspace*{3pt}\cr
    \operatorname{sgn} (\beta) (2-|\beta|)\bu_2\cr
     \qquad{}-\sqrt{1-(2-|\beta|)^2}  \bu_3, &\quad if
    $\beta \in [-2,-1] \cup [1,2]$.\cr
  }
\end{equation}
Figure~\ref{fig:inrha}g and h help to illustrate the
behavior of \eqref{qbeta}, where the location on the ``belt'' is given
by the value of $\beta$.  Note, the diffusion~PDFs have been rotated in
space compared to each other for a better visual perspective.  The
maximum great circle in $q$-space corresponds to the perpendicular
vector $\pm\bu_1$ in space, where the diffusion~PDF exhibits a maximum.
The structure near the peak $(\x=\pm\bu_1)$ is mapped to a structure
contiguous to the great circle, that is, $\q\approx\q(\beta)$.
Comparing the unimodal diffusion models (in Figure~\ref{fig:inrha}a,
b and d), the microstructure of the
diffusion~PDF is mapped into behavior near or on the belt $\q(\beta)$;
see Figure~\ref{fig:inrha}g, h and j.
The scalene structure of the diffusion~PDF corresponds to variation on
the belt (Figure~\ref{fig:inrha}h), while the asymmetry of
Figure \ref{fig:inrha}d and e are mapped onto the local
structure of the delineation of the belt in Figure~\ref{fig:inrha}j
and k.  This motivates us to investigate the structure
of the diffusion~PDF near the great circle of points $\{\q(\beta)\}$
using distances from the great circle to characterize structure in the
decay from the main peak.  To obtain consistency in notation, we define
the set of points, or the great circle perpendicular to $\bu$, via
$\cG(\bu)=\{\q\dvtx\bu^T\q=0,\|\q\|=1\}$ and
$\cG(\bu_1)\equiv\{\q(\beta)\}$.  It is convenient to keep both sets of
notation for ease of exposition in the future.

\section{Scalar summaries and test statistics}

\subsection{Axes of symmetry}

Before we can define appropriate scalar summaries in $q$-space,
additional axes to the $\beta$ axis \eqref{qbeta} are required.  For
any fixed vector $\q(\beta)\in\cG(\bu_1)$ we traverse a great circle
using the vectors
%
\begin{equation}\label{alphabetaplane}
  \q_\perp(\alpha,\beta) = \alpha\bu_1\pm\sqrt{1-\alpha^2}\q(\beta),
  \qquad \alpha \in [-1,1],
\end{equation}
where for $\alpha\in[-2,2]\backslash[-1,1]$, the
corresponding expression may be formed as in \eqref{qbeta}.  Such a
great circle for a fixed value of $\beta$ will be referred to as a
\textit{perpendicular great circle}.

An important component in the definition of our nonparametric summaries
is the \textit{dominant great circle} $\cG(\x_{\max})$ with
$\x_{\max}$ given by
%
\begin{equation}
  \x_{\max}=\arg \max_{\bu}\biggl\{\oint_{\q\in\cG(\bu)}
  \cA(\q)\,d\q \biggr\}.
\end{equation}
If $\cA(\q)$ is an isotropic diffusion process, then $\x_{\max}$ is
any vector in ${\mathbb{R}}^3$ with a fixed norm.  Alternatively, if
${\cA}(\q)$ is ellipsoid with $\lambda_1>\lambda_2\ge\lambda_3$, then
$\x_{\max}=\bu_1$.  If there are two fibers, with relative weights
of $a_1$ and $a_2$ of fiber populations with individual eigenvalues
$\bsL^{(1)}$ and $\bsL^{(2)}$, then
%
\begin{eqnarray}
  \x_{\max} &=& \arg \max_{\bu} \biggl\{ \biggl[ a_1
    \oint_{\q\in\cG(\bu)} \cA_E\bigl(\q; \bsL^{(1)},
    \bsU^{(1)}\bigr)\nonumber\\ [-8pt]\\ [-8pt]
    &&\hphantom{\arg \max_{\bu} \biggl\{ \biggl[}{}+a_2
    \oint_{\q\in\cG(\bu)} \cA_E\bigl(\q; \bsL^{(2)},
    \bsU^{(2)}\bigr)\biggr]\,d\q \biggr\}.\nonumber
\end{eqnarray}
For example, if $a_1\gg{a_2}$, then $\x_{\max}\approx\bu_1^{(1)}$,
or if $a_1=a_2=1/2$ and the great circles do not separate, then
$\x_{\max}$ will lie precisely between the two maxima of the two
diffusion~PDFs.  Once the great circles start to separate the maximum
will go with one of the two.

\subsection{Degree of nonuniformity}

\begin{figure}

\includegraphics{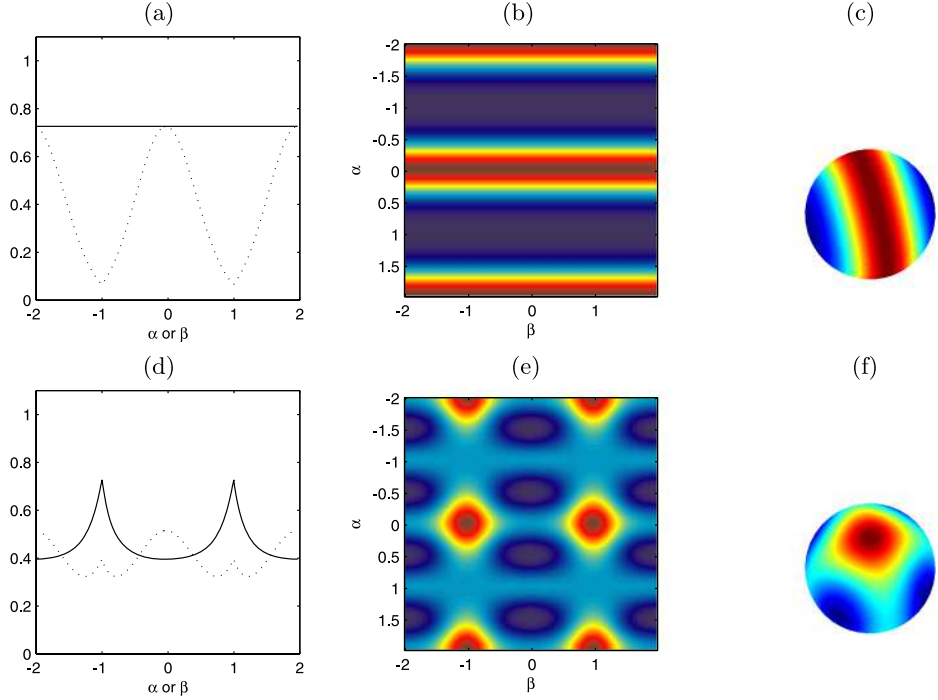}

  \caption{One- and two-dimensional summaries of Gaussian diffusion
    processes in $q$-space, mapped onto the $\alpha$ and $\beta$ axes
    \protect\eqref{alphabetaplane} and their spherical representation.
   \textup{(a, b, c)}  Prolate diffusion process---eigenvalues
    $(\lambda_1\gg\lambda_2=\lambda_3)$.  \textup{(d, e, f)} Mixture of
    two prolate diffusion processes.  The dominant great circle is the
    solid line in the one-dimensional summaries \textup{(a~and d)}, while the
    dotted line is the diffusion from a single perpendicular great
    circle for~\textup{(a)} and the average perpendicular diffusion for \textup{(d)}.
    In the two-dimensional summaries \textup{(b~and~e)} all great circles
    perpendicular to the dominant great circle are plotted on the
    $y$-axis to form the $(\alpha,\beta)$ plane, and the final plots
    in \textup{(c)} and \textup{(f)} show the spherical representation on a single shell
    in Fourier space, corresponding to a fixed wave number magnitude.}
  \label{fig:Gaussian-PDF}
\end{figure}

We represent a unidirectional Gaussian diffusion by plotting the value
of $\cA(\q(\beta))$ (solid line) for $\beta\in[-2,2]$ in
Figure~\ref{fig:Gaussian-PDF}a.  The magnitude on the dominant great
circle is constant over different values of $\beta$ since
$\lambda_2=\lambda_3$.  To illustrate the difference in variation
across the dominant and perpendicular great circles, we also plot the
value of $\cA(\q_{\perp}(\alpha,\beta))$ as a function of $\alpha$ for
a fixed $\beta$ (dotted line).  This line perfectly overlaps
$\cA(\q(\beta))$ at two locations, as it collides with the dominant
great circle when it wraps around the sphere, and decays symmetrically
from~$\q(\beta)$.

We define a new coordinate system $(\alpha,\beta)$, where we expect
consistent variability in $\alpha$ and $\beta$, using our
parameterization of great circles \eqref{alphabetaplane}.  We plot the
unidirectional Gaussian diffusion $\cA(\q_{\perp}(\alpha,\beta))$ for
all perpendicular great circles in the plane
(Figure~\ref{fig:Gaussian-PDF}b).  This prolate diffusion exhibits
variation only in $\alpha$, which is variation perpendicular to the
dominant great circle.  For the prolate diffusion example we can
therefore reduce the variance by averaging across $\beta$ and by
considering the function strictly in terms of $\alpha$.  For comparison
with the $(\alpha,\beta)$ plane, the spherical representation of this
Gaussian diffusion process is provided in
Figure~\ref{fig:Gaussian-PDF}c.

We use the one-dimensional great-circle summaries for a mixture of two
Gaussian diffusions in Figure~\ref{fig:Gaussian-PDF}d, where the
dominant great circle exhibits a large dynamic range relative to the
perpendicular great circles.  In fact, one can determine the number of
peaks of the diffusion~PDF by comparing the dynamic range of the
diffusion between the dominant and perpendicular great circles.  For a
complete picture we also represent the multi-modal diffusion in the
$(\alpha,\beta)$ plane in Figure~\ref{fig:Gaussian-PDF}e, where
variation is appreciable in both the $\alpha$ and $\beta$ axes, and on
the sphere (Figure~\ref{fig:Gaussian-PDF}f).

To overcome the need to compare the variation along the dominant great
circle with all perpendicular great circles individually, we define the
\textit{average perpendicular diffusion} via
%
\begin{equation}\label{e:averagecont}
  \cA_\perp(\alpha) = \frac{1}{2\pi}\int_{0}^{2\pi}
  \cA(\q_\perp(\alpha,\beta(\vartheta)))\,d\vartheta,
\end{equation}
with $\beta(\vartheta)=\cos(\vartheta)$ for $\vartheta\in[0,\pi]$ and
$\beta(\vartheta)=-\cos(\vartheta)-2\mbox{sgn}[\cos(\vartheta)]$
defining $\beta(\vartheta)$ for $\vartheta\in[0,2\pi]$.  One may also
define the average perpendicular diffusion over a half circle by
prespecifying a fixed location on the dominant great circle and
integrating in a window size $\pm1$ around this location.  This will
prevent certain features being masked by the Hermitian symmetry of the
$q$-space measurements.  If $\cA(\q)$ satisfies \eqref{ellipsoid}, then
we have
%
\begin{eqnarray}\label{e:averagecont2}
  \hspace*{25pt}\cA_\perp(\alpha) &=& \frac{1}{2\pi}\int_{0}^{2\pi}
  B\bigl(\bigl(\lambda_1\alpha^2 + (1-\alpha^2)
    [\lambda_2 \|\q(\beta(\vartheta))^T \bu_2\|^2 \nonumber\\ [-8pt]\\ [-8pt]
    &&\phantom{\frac{1}{2\pi}\int_{0}^{2\pi}
  B\bigl((\lambda_1\alpha^2 + (1-\alpha^2)
    [}{}+ \lambda_3
      \|\q(\beta(\vartheta))^T\bu_3\|^2 ]\bigr)^{1/2}\bigr)\,d\vartheta.\nonumber
\end{eqnarray}
Thus, we are averaging the density function over small circles parallel
to the dominant great circle and $\cA_\perp(\alpha)$ measures the
average diffusion at a given value of $\alpha$.  In the special case of
$\lambda_2=\lambda_3$, then
\begin{eqnarray}\label{e:averagecont3}
  \cA_\perp(\alpha) &=& \frac{1}{2\pi} \int_{0}^{2\pi}
  B\bigl(\sqrt{\lambda_1 \alpha^2 + \lambda_2
    [1-\alpha^2]}\bigr)\, d\vartheta\\
  &=& B\bigl(\sqrt{\lambda_1 \alpha^2 + \lambda_2
    [1-\alpha^2]}\bigr).
\end{eqnarray}
The average perpendicular diffusion $\cA_\perp(\alpha)$ provides a
useful summary of variation perpendicular to the dominant great circle.
We define a summary of the diffusion~PDF via
\begin{equation}\label{e:tau}
 \tau = \biggl[\frac{\max_{\alpha}\{\cA_\perp(\alpha)\}}
   {\min_{\alpha}\{\cA_\perp(\alpha)\}}\biggr] \Big/
   \biggl[\frac{\max_{\beta}\{\cA(\q_{\perp}(0,\beta))\}}
     {\min_{\beta}\{\cA(\q_{\perp}(0,\beta))\}}\biggr] - 1.
\end{equation}
If the diffusion is isotropic, we know that
$\lambda_1=\lambda_2=\lambda_3$.  In this case we have
${\cA}_\perp(\alpha_{\max})={\cA}_\perp(\alpha_{\min}) =
B(\sqrt{\lambda_1})$ and $\cA(\q(0,\beta_{\max})) =
\cA(\q(0,\beta_{\min}))=B(\sqrt{\lambda_1})$, resulting
in $\tau=0$.  If the diffusion is ellipsoidal and
$\lambda_2=\lambda_3$, then
$\tau=B(\sqrt{\lambda_2})/B(\sqrt{\lambda_1})-1>0$.
If we adopt the mixture model, with multiple peaks, then it is possible
to get $\tau\gg0$ even if we do not have a single diffusion~PDF and we
define
\begin{equation}
  \tilde{\tau} = \min_{\beta} \max_{\alpha_1,\alpha_2}
  \biggl\{\frac{\cA(\q_\perp(\alpha_1,\beta))}
  {\cA(\q_\perp(\alpha_2,\beta))}\biggr\} \Big/
  \biggl[\frac{\cA(\q_{\perp}(0,\beta_{\max}))}
  {\cA(\q_{\perp}(0,\beta_{\min}))}\biggr] - 1.
    \label{tautilde}
\end{equation}
We note that under isotropy $\tilde\tau\equiv0$, while if we have a
single ellipsoid diffusion $\tilde\tau\equiv\tau>0$.  For a double
tensor model $\tilde\tau$ is more robust and will (in general) take on
a lower value compared with $\tau$.  In contrast to $\tau$ and
$\tilde{\tau}$, we could also study the variability in the $q$-space
density directly in terms of the ODF.  \cite{Tuch}, for example,
defines the generalized fractional anisotropy (GFA) via
\begin{equation}
  \mbox{GFA} = \biggl\{\frac{n\sum_{i=1}^{n}
    (\ODF_\mathrm{W}(\theta_i,\phi_i) - 1/n)^2}
       {(n-1)\sum_{i=1}^n\ODF_\mathrm{W}^2(\theta_i,\phi_i)}\biggr\}^{1/2},
\end{equation}
and this measures the nonuniformity of the spatial distribution, as do
also the normalized entropy and the nematic order parameter
[\cite{Tuch}].  While the GFA quantifies the lack of uniformity in the
ODF, if there is more than one fiber, determining its statistical
properties is nontrivial, unlike the case for $\tau$ and $\tilde\tau$.
Another such measure, generalized anisotropy is defined in terms of the
generalized trace of the tensor representation of the mean diffusivity
[\cite{Ozarslan2005}].

\subsection{Measures of anisotropy}

To determine the importance of the identified dominant great circle (or
orientation), we can, with a model of \eqref{ellipsoid}, compare
$B(\sqrt{\lambda_1})$ to $B(\sqrt{\lambda_2})$ and
$B(\sqrt{\lambda_3})$.  We define the following \textit{anisotropy
  statistic} to perform such a comparison:
\begin{equation}\label{e:anisotropy}
  \xi = \frac{\log [{\cA}_\perp(0)]}{\log [{\cA}_\perp(1)]} =
  \frac{\log [B(\sqrt{\lambda_2})]}
       {\log [B(\sqrt{\lambda_1})]},
\end{equation}
where the last equality follows if $\lambda_3=\lambda_2$.  This
statistic measures the degree of anisotropy over the $q$-space shell by
comparing the peak-to-trough values (i.e., the value at the maximum
great circle, compared to the value at the single point perpendicular
to that maximum).  Figure~\ref{fig:Gaussian-PDF}a displays the
difference between the maximum and minimum for an average perpendicular
great circle.

\begin{figure}

\includegraphics{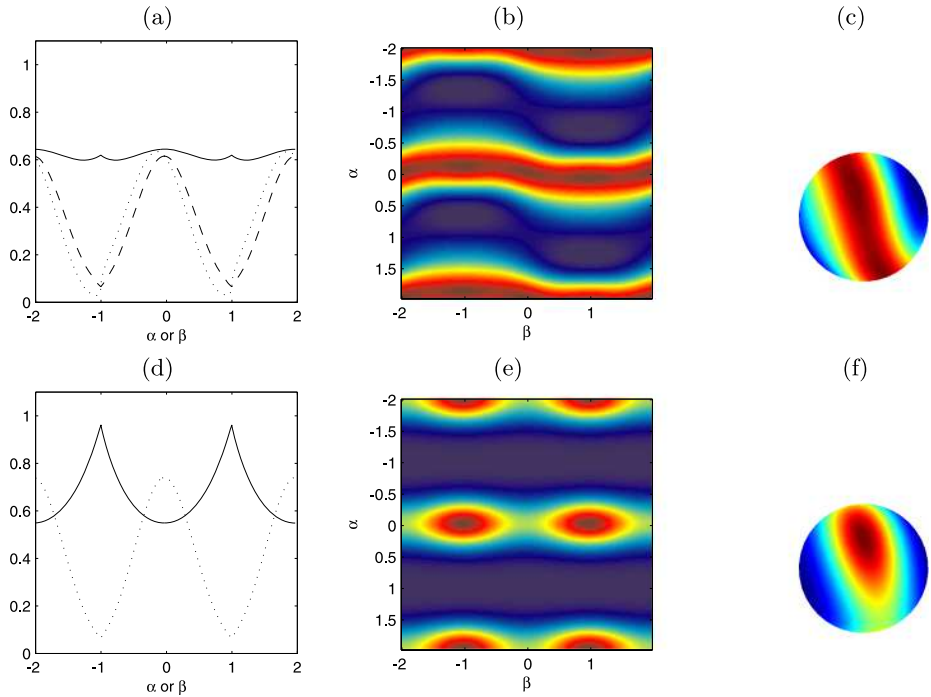}

  \caption{One- and two-dimensional summaries of Gaussian diffusion
    processes in $q$-space, mapped out in the $\alpha$ and $\beta$
    axes \protect\eqref{alphabetaplane} and their spherical representation.
    \textup{(a, b, c)}  An asymmetric diffusion process.  This is apparent
    by the asymmetric decay in great circles perpendicular to the
    dominant great circle in the $(\alpha,\beta)$ plane.
    \textup{(d, e, f)} A scalene diffusion process with eigenvalues
    $(\lambda_1\approx\lambda_2\gg\lambda_3)$.  The dominant great
    circle is the solid line in the one-dimensional summaries \textup{(a and
    d)}, while the dotted line is the the average perpendicular
    diffusion over $\beta\in[-2,0]$ for \textup{(a)} and all $\beta$'s for \textup{(d)}.
    The dashed line in \textup{(a)} gives the average over all $\beta$'s.  In
    the two-dimensional summaries all great circles perpendicular to
    the dominant great circle are plotted on the $y$-axis to form the
    $(\alpha,\beta)$ plane.  The final plots in \textup{(c)} and \textup{(f)} show the
    spherical representation on a single shell in Fourier space,
    corresponding to a fixed wave number magnitude.}
  \label{fig:Gaussian-PDF2}
\end{figure}

The \textit{decay ratio statistic} quantifies the variability of the
diffusion over the dominant great circle
\begin{equation}\label{zieq}
  \zeta = \max_{\beta}\frac{\log[\cA(\q(\beta))]}{\log[\cA(\q(\beta+1))]}.
\end{equation}
When the two smaller eigenvalues ($\lambda_2$ and $\lambda_3$) are
approximately equal then $\zeta\approx1$, otherwise $\zeta\gg1$. The
scalene diffusion in Figure~\ref{fig:Gaussian-PDF2}c
and d exhibits such structure $(\zeta\gg1)$.

An indication of forking in white matter would correspond to an
asymmetric decay of the diffusion~PDF associated with different decays
depending on the parity of the deviation.  In this case we may no
longer model the diffusion~PDF as ellipsoidal.  For example, in
Figure~\ref{fig:Gaussian-PDF2}a and b we see
that while there is still a strong orientation from the dominant great
circle, the PDF no longer exhibits symmetric decay away from the
dominant great circle.  Note that the decay is symmetric in $\alpha$
when averaged over the full sphere to produce ${\cA}_{\perp}(\alpha)$.
Hence, averaging over $\beta$ is not appropriate if we want to detect
asymmetry since a symmetric distribution will be obtained from the
Hermitian symmetry of the HARDI measurements when averaging over a full
great circle.  A suitable \textit{asymmetry statistic} to measure
potential asymmetry is given by
\begin{eqnarray}
  \kappa(\beta) &=& \frac{(1/2)\int_{0}^{\pi/2}
    [\cA(\q_\perp(\alpha(\vartheta),\beta)) -
      \cA(\q_\perp(-\alpha(\vartheta),\beta))]\,d\vartheta}
        { \int_{0}^{\pi/2}
          \cA(\q_\perp(\alpha(\vartheta),\beta))\,d\vartheta},\\
        \vartheta_{\max} &=& \arg \max\kappa(\beta(\vartheta)),
        \qquad \beta_{\max}=\beta( \vartheta_{\max}),\\\label{eq:kappa}
        \kappa &=& \frac{2}{\pi}\int_{\vartheta_{\max}-\pi/4}^{\vartheta_{\max}+\pi/4}
        \kappa(\beta(\vartheta))\,d\vartheta.
\end{eqnarray}
The definition of $\kappa$ is motivated by the wish to both obtain a
test statistic with sufficient power and also to reduce its variance.
The discrete approximation to $\kappa$ will have a smaller variance
than $\kappa(\beta_{\max})$.  Asymmetry in the decay from the main
peak may occur when the PDF is a mixture of diffusions with varying
strengths.  If the two populations are sufficiently separated and
equivalent in magnitude, then this will be indicated by $\tau$ and/or
$\tilde\tau$ and the diffusion will be recognized as a so-called
``crossing fiber.''  If the mixture of diffusions contains two
different strengths, then the dominating PDF will be recognized when
determining $\x_{\max}$.  The remaining structure will not be fully
consistent with a single tensor and will (in general) appear to be
asymmetric compared to the dominant great circle.

\begin{table}
  \caption{The structure of the proposed diffusion models and the
    representation of their structure in terms of the proposed
    statistical summaries.  Key to abbreviation where the statistics
    represent N-P$/$A (non-preference versus anisotropy), C$/$E (circular
    versus ellipsoidal), S$/$A (symmetric versus asymmetric), I$/$M
    (isotropic versus multi-modal), M$/$U (multi-modal versus unimodal)}
  \label{tab:summaryofprop}
  \begin{tabular*}{\textwidth}{@{\extracolsep{\fill}}lcccccc@{}}
    \hline
    \textbf{Hypothesis} & \textbf{Statistic} & \textbf{Isotropic} & \textbf{Prolate}
    & \textbf{Scalene} & \textbf{Mixture} & \textbf{Heterogeneous}\\
    \hline
    N-P$/$A & $\tau$ & small & large & large & small & large \\
    M$/$U & $\tilde\tau$ & small & large & large & small & large \\
    I$/$M & $\xi$ & one  & small & small & small & small\\
    C$/$E & $\zeta$& -- & one & large & -- & large\\
    S$/$A & $\kappa$& -- & zero & zero & -- & large\\
    \hline
  \end{tabular*}
\end{table}

Let us discuss models that will lead to a different structure in the
proposed summaries.  We refer to Table~\ref{tab:summaryofprop} to
summarize the properties of each statistical test, and different
diffusion~PDFs lead to different structures.  It may seem insufficient
to consider only an isotropic PDF, a single peak, a~double peak, or
something more heterogeneous.  However, even with a fully parametric
model of a Gaussian DTI framework, a two-tensor model has 13
(identifiable) parameters and a three-tensor model has 19.  If one
considers acquiring 60 gradient encoding directions (a common sample
size), then one is forced to fit a highly-saturated model that results
in noisy estimates---especially at higher $b$-values where the
orientational heterogeneity can be well resolved.  Pushing much beyond
a small number of parameters or features of interest is not advisable
with such sampling.

\section{Estimation}
\label{matmeth}

\subsection{Parameterizing the $(\alpha,\beta)$ axes}

Having proposed various summaries of the population of diffusion~PDFs
at a particular voxel, these must now be estimated from a set of
diffusion measurements.  The dominant direction may be estimated via
\begin{eqnarray}\label{eq:lala}
  \hat{\bu}_{1} &=& \x_{\max} =
  \arg\max_{\bu,\|\bu\|=1}\biggl\{ \int_{\q\in
  \cG(\bu)}\wh{\cA}(\q)\,d\q \biggr\}\nonumber\\ [-8pt]\\ [-8pt]
  &\equiv& \arg
  \max_{\x} \operatorname{FRT}\{\wh{\cA}\}(\x),\nonumber
\end{eqnarray}
where $\operatorname{FRT}\{\cdot\}$ denotes the Funk--Radon Transform (FRT) as
utilized in \cite{Tuch}.  Note that $\wh\cA(\q)$ refers to the
multiresolution-based estimator [\cite{OlhedeWhitcher} and may be
replaced by another appropriate estimator.  We assume the availability
of the quantity $(\hat\sigma^{\ast})^2$, an estimator of the variance
of the error in $A(\q_k)$ which we define to be $\sigma^2$.  The
variance of $\wh\cA(\q_k)$ is assumed to be $\tilde\sigma^2\le\sigma^2$
and the variance of an interpolated value of the diffusion~PDF is
$\breve\sigma^2\le\tilde\sigma^2\le\sigma^2$. The integral may be
approximated numerically by interpolating the observed HARDI
measurements at evenly-spaced points along several great circles, each
perpendicular to a given $\x_i$.

The effects of using different numerical methods for this step is a
trade-off between increasing numerical accuracy and decreasing
variance.  Interpolating using spherical harmonics reduces variance but
may smooth out details depending on the choice of regularization; see
the discussion in \cite{Descoteaux} and \cite{Hess}.  We instead use
a locally linear interpolation on the polar representation of the
observed data that enforces the periodicity of the space in which it
was sampled.  Simple structures in terms of the observed points can mix
over several spherical harmonics, and so the magnitude of individual
spherical harmonic coefficients may not be large, even if the local
coefficient is large.  This fact makes the representation inappropriate
for using the smoothing methods of previous authors. The choice of
interpolation procedure should be considered in terms of which
statistic one is using, as the variance and bias must be balanced
specifically for this purpose.  We also note that spherical harmonics
do not possess the same properties as Fourier vectors, and that an
infinite number of harmonics are required for perfect reconstruction of
a surface on a sphere, and so any reconstruction from the continuous
basis will be inaccurate.

The spatial maximum is determined from $\{\operatorname{FRT}\{\cA\}(\x_i)\}_i$.
The spatial location $\x_{\max}$ is an estimator of $\bu_1$ and we
estimate a vector in the linear subspace spanned by $\bu_2$ and $\bu_3$
from the eigensystem of $\bld{I}-\x_{\max}\x_{\max}^T$, this
yielding $\hat{\bu}_2$ and $\hat{\bu}_3$, that maximize the difference in
decay in the two axes. For numerical implementation we sample the
estimated dominant great circle by discretizing $\alpha$ to
$\{\alpha_j\}_{j=1}^N$ and $\beta$ to
$\{\beta_k\}_{k=1}^N$, for an even integer $N$.  A
discretized version of \eqref{qbeta} is then given by
\begin{equation}\label{e:discgreatcircle}
 \hspace*{15pt}\hat\q_k = \cases{
  (2-\beta_k) \hat{\bu}_2\vspace*{2pt}\cr
  \qquad{}-\sqrt{1-(2-\beta_k)^2}\hat{\bu}_3,
    &\quad for  $k=-N/4,\dots,-1$,\cr
    \beta_k \hat{\bu}_2 + \sqrt{1-\beta_k^2}\hat{\bu}_3,
    &\quad for  $k=0,\dots,N/2-1$,\vspace*{2pt}\cr
    (-2-\beta_k) \hat{\bu}_2\cr
    \qquad{} -\sqrt{1-(2+\beta_k)^2}\hat{\bu}_3,
    &\quad for $k=N/2,\dots,3N/4-1$,}
\end{equation}
where
\begin{equation}
  \beta_k = \cases{
    2 - \cos(2\pi k/N), &\quad for $k=-N/4,\dots,-1$,\cr
    \cos(2\pi k/N), &\quad for  $k=0,\dots,N/2-1$,\cr
    -2 - \cos(2\pi k/N), &\quad for  $k=N/2,\dots,3N/4-1$.
 }
\end{equation}
We define $\hat\q_k$ for the values of $k$ \textit{not} between
$k=-N/4,\dots,3N/4-1$ by cyclically extending
\eqref{e:discgreatcircle}.  The choice of discretization guarantees the
distance between the orientation associated with a great circle and the
great circle is one.  The parameters $\alpha_j$ and $\beta_j$ are
individually discretized to force equal length increments on the great
circle.  Thus, we discretize the estimated dominant great circle via
$\{\hat\q_k\}$.  Once $\x_{\max}$ has been determined, the
diffusion may be characterized directly in $q$-space.  Let us define
the sampled great circle vectors for $\{\hat\q_k\}$ in
\eqref{e:discgreatcircle} via
\begin{equation}\label{e:discperpgreatcircle}
  \hspace*{15pt}\hat\q_{\perp jk} = \cases{
    (2-\alpha_j) \hat{\bu}_{1}\vspace*{2pt}\cr
    \qquad{} - \sqrt{1-(2-\alpha_j)^2}\hat{\q}_{k},
    &\quad for $j=-N/4,\dots,-1$,\cr
    \alpha_j \hat{\bu}_{1} + \sqrt{1-\alpha_j^2}\hat{\q}_{k},
    &\quad for  $j=0,\dots,N/2-1$,\vspace*{2pt}\cr
    (-2-\alpha_j) \hat{\bu}_{1}\vspace*{2pt}\cr
    \qquad{} - \sqrt{1-(2+\alpha_j)^2}\hat{\q}_{k},
    &\quad for $j=N/2,\dots,3N/4-1$;
 }
\end{equation}
where
\begin{equation}
  \alpha_j = \cases{
    2 - \cos(2\pi j/N), &\quad for  $j=-N/4,\dots,-1$,\cr
    \cos(2\pi j/N), &\quad for  $j=0,\dots,N/2-1$,\cr
    -2 - \cos(2\pi j/N), &\quad for  $j=N/2,\dots,3N/4-1$.
  }
\end{equation}
A discretized version of the average perpendicular diffusion
\eqref{e:averagecont} is given by
$\wh{\cA}_\perp(\alpha_j)=N^{-1}\sum_{k}\wh{\cA}(\hat\q_{\perp jk})$,
and we can sum over any $N$ consecutive $k$ (e.g., it does not matter
exactly how we sum over $k$) because of the periodic extension.

\subsection{Diagnosing nonuniformity}

In order to test large-scale properties of the diffusion directly in
$q$-space, we consider the following statistical hypothesis
$H_0\dvtx\cA(\q)=\cA~\forall\q$ versus $H_1\dvtx\cA(\q)=\cA_\mathrm{E}(\q)$.  Our test
statistic is based on a discretized version of \eqref{e:tau}, given by
\begin{equation}
  T = \biggl[\frac{\max_j\{\wh{\cA}_\perp(\alpha_{j})\}}
    {\min_j\{\cA_\perp(\alpha_{j})\}}\biggr] \Big/
  \biggl[\frac{\max_k\{\cA(\hat{\q}_{ k})\}}
    {\min_k\{\cA(\hat{\q}_{ k})\}}\biggr] - 1.
\end{equation}
The distribution of this test statistic is derived in \hyperref[suppA]{Supplementary Material}.  If
the observations are isotropic, then the properties along the dominant
great circle will be equivalent to the properties on the perpendicular
great circle (ignoring any random/discretization errors).  The
estimators of $\cA$ and~$\tilde\sigma$, under the null of
$\wh{\cA}(\hat\q_k)\cong\cA+\tilde\sigma\epsilon$, are given by
\begin{eqnarray}\label{e:estunderh0}
  \ol{\cA}_N &=& \frac{1}{N} \sum_{k=1}^N \wh{\cA}(\hat\q_k)
  \stackrel{N}{\rightarrow}
  \frac{1}{2\pi}\int_{\cG(\hat{\bm{\upsilon}}_1)}\cA(\q)\,d\q,\\
  \hat\sigma_{\cA} &=& \sqrt{\rho}
  \operatorname{MAD}\{\wh{\cA}(\hat\q_k)-\ol{\cA}_N\dvtx k=1,\dots,N\},
\end{eqnarray}
where $0<\rho\le{3}$, $\hat\q_k$ defined by \eqref{e:discgreatcircle}
and $\operatorname{MAD}\{\cdot\}$ is the maximum absolute deviation.  These
equations provide estimators of the mean value of the isotropic
diffusion and the standard deviation of $\wh{\cA}(\q)$ at the observed
measurements.  The parameter $\rho$ is a constant depending on the
linear interpolation method used for the implementation.  Taking a
value of $\rho=3$ is suitable for our choice of numerical interpolation
and we define
\begin{equation}\label{e:Um}
  U = T \frac{\overline{\cA}_N}{\hat\sigma_{\cA}}.
\end{equation}
We can compute the critical value $u_{\alpha}$ using the fact that
$F_U(u_{\alpha})=1-\alpha$, where $F_U(\cdot)$ is defined in
\hyperref[suppA]{Supplementary Material}.  We report two critical values here, $u_{0.05}=0.1185$ for
the $m$ which is consistent with our sampling scheme, and
$u_{0.05}^{(\mathrm{con})}=1.9637$ with a conservative distribution
approximation.

We also develop a new test based on a null of a multi-modal diffusion,
where we define multi-modal in terms of
$(\wt{\cA}_{{\max}}\cA_{\min})/(\wt{\cA}_{{\min}}\cA_{\max})<c=2$,
where $\wt{\cA}_{{\max}}$ and $\wt{\cA}_{{\min}}$ are the
maximum and minimum on the perpendicular great circle minimizing~\eqref{tautilde} in $\beta$, respectively, while $\cA_{{\max}}$ and
$\cA_{{\min}}$ are the maximum and minimum on the dominant great
circle.  The value\vadjust{\goodbreak} of $c$ is fairly arbitrary, but to develop a
powerful method of separating the clearly unimodal from the
multi-modal, some level must be chosen based on the deterministic
structure of the sampled diffusion~PDF.  To separate unimodal from
multi-modal PDFs, we start from $\tilde\tau$ and define
\begin{equation}
  \wt{T} = \min_{k} \biggl\{\max_{{j_1},{j_2}}
  \biggl\{\frac{\wh{\cA}(\hat\q_{\perp j_1 k})}
              {\wh{\cA}(\hat\q_{\perp j_2 k})}\biggr\}\biggr\}
              \Big/ \biggl[\frac{\max_k\{\wh{\cA}(\hat\q_{k})\}}
                {\min_k\{\cA(\hat\q_{k})\}}\biggr] - 1.
\end{equation}
We shall now choose to distinguish the multi-modal from the unimodal,
and so normalize the statistic using
$\wt{U}=(\wt{T}-[c-1])\wh{\cA}_{{\min}}/(\hat\sigma_{\cA}\sqrt{2c^2+2})$,
where $\wh{\cA}_{{\min}}=\wh{\cA}(\wh{\bm{\upsilon}}_1)$.  The
distribution of this test statistic is derived in \hyperref[suppA]{Supplementary Material}, under
the specified null hypothesis.

If, on the other hand, we have failed to reject the null hypothesis
``$\cA(\q_{\perp}(\alpha,\break\beta))$ equally variable in $\beta$ for
$\alpha=0$ as it is in $\alpha$ for a fixed $\beta$,'' then based on
the $T$-statistic we need to distinguish voxels that indicate two fiber
populations versus isotropic voxels.  Let us define a discrete version
of \eqref{e:anisotropy} to be
\begin{equation}\label{e:degreeofanisotropy}
  X = \frac{\log[\wh{\cA}_\perp(0)]}{\log[\wh{\cA}_\perp(1)]}.
\end{equation}
We can interpret $\xi$, and the sample version $X$, as the degree of
anisotropy from the average perpendicular great circle.  We recognize
that the statistic~$X$ is comparing the average apparent diffusion
coefficient (ADC) on the great circle to the average ADC perpendicular
to the great circle, or that~\eqref{e:degreeofanisotropy} may be
rewritten in terms of the ADC at a fixed value of $b$ via
%
\begin{equation}
  X = \sum_k\wh{C}(\hat\q_k)\Big/\sum_k\wh{C}(\hat\q_{\perp N/4 k}).
\end{equation}
The ADC is $\wh{C}(\tilde\q_j)=-b^{-1}\log{A(\tilde\q_j)}$, for a
defined set of $\q$ vectors $\tilde{\q}_j$
[\cite{alebararrdetection}].  With an assumption of ellipsoidal
structure~\eqref{ellipsoid} we have averaged the ADC to reduce variance
when estimating~$\xi$ without accruing bias.  We define $X_k$ as the
sample anisotropy calculated using only the $k$th perpendicular great
circle via
\begin{equation}
  X_k = \frac{\log[\wh{\cA}(\hat\q_{k})]}{\log[\wh{\cA}(\hat\q_{\perp{N/4}k})]},
\end{equation}
and refer to \eqref{e:discperpgreatcircle} for the definition of
$\hat\q_{\perp jk}$.  Under moderate-to-high SNR~$X_k$ may be
approximated by a Gaussian random variable.  We quantify uncertainty,
when there are potentially several peaks, using
$\hat\sigma_2=\min\{\hat\sigma_{\cA},\hat\sigma^{\ast}\}$, where
$\hat\sigma_{\cA}$ is defined in \eqref{e:estunderh0} and
$\hat\sigma^{\ast}$ is the available estimator for $\sigma$.  By using
the minimum, we ensure that the estimated variance is not inflated
compared to its pre-smoothing value.

For those voxels where isotropy cannot be rejected, we may now
distinguish between isotropy and a multiple-tensor model using $X$. The
\textit{multi-modality} statistic is given by
\begin{equation}
 Q = \frac{\rho(X-1)}{\hat\sigma_{2}}
  |\overline{\cA}_N\log\overline{\cA}_N|.
\end{equation}
So we consider the test $H_0\dvtx\cA(\q)=\cA~\forall\q$ versus
$H_1\dvtx\max_k\{\cA(\q_{k})\}\gg\break\min_k\{\cA(\q_{k})\}$ (multiple peaks)
and use $Q$ as the test statistic, whose distribution under the null is
provided in \hyperref[suppA]{Supplementary Material}.  The three tests outlined here allow one to at
a single voxel diagnose the structure of the diffusion~PDF, where $U$
is used to separate anisotropic~PDFs from isotropic~PDFs, $\wt{U}$ is
used to separate ellipsoid~PDFs from multi-modal~PDFs and $Q$ is used
to separate multi-modal~PDFs from isotropic~PDFs.

\subsection{Diagnosing asymmetry}

Having established methodology to discriminate the number of peaks in
the diffusion~PDF at a single voxel, we now provide additional
methodology to characterize the diffusion~PDF as scalene versus other
forms of asymmetry, for example, to observe the indication of forking
or fanning white-matter structure.  Let us define
\begin{equation}
  k_{\max} = \arg\max_{1\le k\le N/4}
  \frac{\log\wh{\cA}(\hat{\q}_k)}{\log\wh{\cA}(\hat{\q}_{k+N/4})}.
\end{equation}
The effective degrees of freedom parameters $(m,m')$ are related via
$m<m'<2m$, for robustness, so that
\begin{equation}\label{defofZ}
  Z =\frac{\log\wh{\cA}(\hat{\q}_{{k}_{\max}+N/(2m')})}
  {\log\wh\cA(\hat{\q}_{{k}_{\max}+N/(2m')+N/4})}.
\end{equation}
We remark that $Z$ is related in some sense to $X_k$ (refer to
Figures~\ref{fig:Gaussian-PDF} and~\ref{fig:Gaussian-PDF2}).  The
statistic $X_k$ compares the value of the diffusion~PDF at location $k$
on the dominant great circle $(\alpha=0)$ to the value at the
perpendicular to the dominant great circle $(\alpha\neq0)$.  The
statistic $Z$ in contrast looks at the difference in values of the
diffusion~PDF on the great circle itself ($\alpha=0$ and~$\beta$
varies).  Under the null hypothesis $\cA(\cdot)$ is constant on the
great circle, and if the medium and minor eigenvalues are approximately
equal, then $\E\{Z\}=\zeta\approx{1}$, otherwise $\zeta\gg1$.  A
normalized version of the decay ratio statistic \eqref{defofZ} is given
by
\begin{equation}
  V = \frac{(Z-1)|\overline{\cA}_N\log\overline{\cA}_N|}%
  {\hat\sigma_2}.
\end{equation}
A suitable threshold for this statistic may be found in \hyperref[suppA]{Supplementary Material}.
The statistics $Q$ and $V$, used to test different hypotheses of
nonisotropic decay, have similar forms.

The summary statistic $\kappa$ in \eqref{eq:kappa} allows one to
diagnose white-matter microstructure that is not consistent with a
single ellipsoid diffusion.  Departures from such a single ellipsoid
diffusion model may be attributed to partial-volume effects, or a
heterogeneous population of fibers [\cite{Behrens2007}].  For such a
model \eqref{ellipsoid} is no longer appropriate and we would rather
fit a mixture model with unequal populations---or possibly
$\cA_{\mathrm{DE}}(\cdot)$.  In such circumstances one cannot use the average
perpendicular great circle to uncover asymmetry since averaging over
all possible $\beta$'s will produce a symmetric distribution regardless
of the underlying fiber characteristics.  Taking
$\breve{k}_{\max}=\arg\max_k\{P_k\}$, we define the
\textit{asymmetry statistic} via
\begin{eqnarray}
  K &=& \frac{1}{N/4+1}
  \sum_{k=\breve{k}_{\max}-N/8}^{\breve{k}_{\max}+N/8} P_k,\\
  P_k &=& \frac{8\sum_{j=1}^{N/4-1} [\wh{\cA}(\hat\q_{\perp jk}) -
      \wh{\cA}(\hat\q_{\perp(j+N/4)k})]} {\sum_{j=1}^N
    \wh{\cA}(\hat\q_{\perp jk})}.
\end{eqnarray}
Full details on the distribution of $K$ may be found in \hyperref[suppA]{Supplementary Material}.
We~have chosen $N/8$ to improve the power, since averaging decreases the
variance, but the asymmetry is greatest near the maximum (compare with
Figure \ref{fig:Gaussian-PDF2}b).  For tests at a specific voxel we
perform the hypothesis test $H_0\dvtx\kappa=0$ versus $H_1\dvtx\kappa\neq0$,
using quantiles from the standard Gaussian~PDF $\phi(\cdot)$.  This
text identifies diffusion~PDFs that are non-Gaussian in terms of the
parity structure in the principal axes. However, it does not compare
the maximum and minimum of a perpendicular great circle, rather it
finds a set of perpendicular great circles for which the decay around
the dominant great circle is asymmetric and estimates this average
asymmetry, for example, Figure~\ref{fig:Gaussian-PDF2}b.

The usage of the statistics is now combined at a voxel level.  The most
important step is to classify the voxel as isotropic, unimodal or
multi-modal.  With this information the local structure of the peaks in
the diffusion~PDF may be further characterized.  This is similar to the
situation when comparing PDFs between voxels for fiber tracking
(tractography) where the number of mixture components in the voxels is
the first priority and then the components in the diffusion~PDF are
matched between voxels using local characteristics corresponding to
structures at fine scales.  From knowledge of white-matter structure
one would anticipate varying values of asymmetry before a forking fiber
structure, and this would allow us to smoothly go between single voxels
with unimodal diffusion to mixtures.  These topics shall be discussed
in subsequent sections.

\begin{figure}[b]

\includegraphics{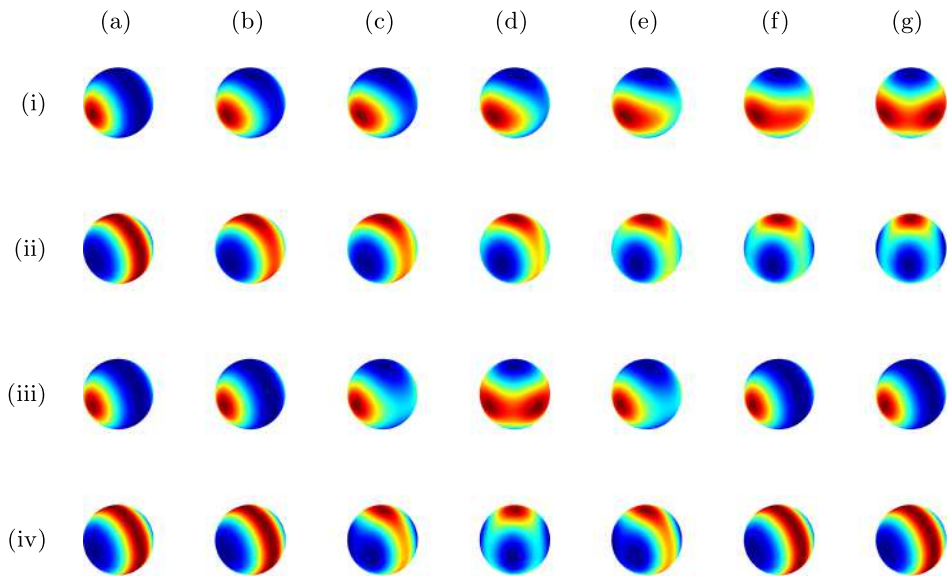}

  \caption{An illustration of the evolution of a diffusion~PDF through
    a number of adjacent voxels in space and $q$-space.  The first and
    second rows are the spatial and $q$-space evolution, respectively,
    of the diffusion~PDF for a forking fiber.  The third and fourth
    rows are the spatial and $q$-space evolution, respectively, of the
    diffusion~PDF for a crossing fiber.  The aim of the plots is to
    show the changing $q$-space structure over this evolution.}
  \label{fig:evolution}
\end{figure}

\subsection{Example: Crossing and forking fibers}

We consider two typical heterogeneous white-matter structures, a
crossing fiber and a forking fiber (Figure~\ref{fig:evolution}).  The
spatial representation of the forking fiber is provided in the first
row of the figure, denoted by (i), while the $q$-space representation
of the forking fiber is given on the second row of the figure, denoted
by (ii).  In the spatial representation we see a~single fiber
population in voxel~(i,a) and, as we traverse from left-to-right, the
two populations become more apparent until a second fiber appears in
voxel~(i,g).  The $q$-space version of these two populations shows a
scalene distribution developing in voxels~(ii,b)--(ii,e).  As the
forking progresses, from left-to-right, it appears both highly warped
and scalene until the distribution\vadjust{\goodbreak} clearly displays multiple fibers in
voxel~(iii,g).  These fibers were generated by aggregating two
densities via
\begin{eqnarray}\label{example-fibers}
  \cA(\q,t) &=& a_1(t) \cA_g \bigl(\q; \bsL^{(1)}(t),
  \bsU^{(1)}(t)\bigr)\nonumber\\ [-8pt]\\ [-8pt]
  &&{}+ \bigl(1-a_1(t)\bigr) \cA_g\bigl(\q; \bsL^{(2)}(t), \bsU^{(2)}(t)\bigr).\nonumber
\end{eqnarray}
In the case of the forking fiber, $a_1(t)=1-t/2$ and the main
directions of $\bsU^{(1)}(t)$ and $\bsU^{(2)}(t)$ are given by
$(1,0,0)$ and $(\cos(\pi{t}/2),\sin(\pi{t}/2),0)$, respectively.
The individual tensors take values similar to $\cA_1(\q)$ and
$t\in[0,1]$.  The spatial representation of the crossing fiber is the
third row (iii) of Figure~\ref{fig:evolution}, with its corresponding
$q$-space representation in the fourth row~(iv).  The ellipsoid appears
prolate in voxel~(iii,a), then two fiber populations are present in
voxel~(iii,d) and eventually the fiber population returns to a~prolate
shape.  With respect to the parameterization of the crossing fiber in
\eqref{example-fibers}, $a_1(t)\in\{1,0.75,0.5\}$ and the two fibers
cross at 90~degrees with parameters similar to~$\cA_1(\q)$.

\begin{table}[b]
  \caption{Discretized summaries based on nonparametric measures of
    symmetry for a modeled forking and crossing fiber, compare with
    Table~\protect\ref{tab:summaryofprop}.  The summaries show how the
    statistics evolve over a sequence of voxels undergoing forking or
    crossing}
  \label{raw_properties}
  \begin{tabular*}{\textwidth}{@{\extracolsep{\fill}}lccccccc@{}}
    \hline
     & \multicolumn{7}{c@{}}{\textbf{Forking fiber}}\\ [-7pt]
   & \multicolumn{7}{c@{}}{\hrulefill}\\
    \textbf{Statistic}  & \textbf{(i,a)} & \textbf{(i,b)} & \textbf{(i,c)}
    & \textbf{(i,d)} & \textbf{(i,e)} & \textbf{(i,f)} & \textbf{(i,g)}\\
    \hline
    $\tau$           & \phantom{.}10.18  & 5.12 & 3.35 & \phantom{$-$}1.86 & 0.82 & 0.07 & $-$0.14\\
    $\tilde{\tau}$   & \phantom{$-$}8.98  & 4.94 & 3.36 & \phantom{$-$}2.13 & 1.19 & 0.33 & $-$0.07\\
    $\xi$            & \phantom{$-$}0.12  & 0.18 & 0.21 & \phantom{$-$}0.27 & 0.35 & 0.53  & \phantom{$-$}0.69\\
    $\zeta$          & \phantom{$-$}1.03  & 1.51 & 1.72 & \phantom{$-$}1.91 & 2.06 & 2.21  & \phantom{$-$}2.67\\
    $\kappa$         & $-$0.01 & 0.03 & 0.17 & \phantom{$-$}0.35 & 0.45 & 0.54  &
    \phantom{$-$}0.30\\ [6pt]
            & \multicolumn{7}{c@{}}{\textbf{Crossing fiber}}\\ [-7pt]
         & \multicolumn{7}{c@{}}{\hrulefill}\\
     \textbf{Statistic}  & \textbf{(iii,a)} & \textbf{(iii,b)} & \textbf{(iii,c)}
     & \textbf{(iii,d)} & \textbf{(iii,e)} & \textbf{(iii,f)}\\
    \hline
    $\tau$           & \phantom{$-$}9.19 & 9.19 & 1.15 & $-$0.14 & 1.15 & 9.19\\
    $\tilde\tau$     & \phantom{$-$}8.98 & 8.98 & 1.26 & $-$0.07 & 1.26 & 8.98\\
    $\xi$            & \phantom{$-$}0.12 & 0.12 & 0.32 &  \phantom{$-$}0.69 & 0.32 & 0.12\\
    $\zeta$          & \phantom{$-$}1.04 & 1.04 & 1.74 &  \phantom{$-$}2.67 & 1.74 & 1.04\\
    $\kappa$         & \phantom{$-$}0.00 & 0.00 & 0.09 &  \phantom{$-$}0.30 & 0.09 & 0.00\\
    \hline
  \end{tabular*}
\end{table}

The description of a crossing fiber is in many ways simpler than a
forking fiber.  Table~\ref{raw_properties} lists the summary statistics
$(\tau,\tilde\tau,\xi,\zeta,\kappa)$ for the crossing and forking fiber
examples in Figure~\ref{fig:evolution}.  Note that these deterministic
summaries have not been normalized, unlike the statistics in
Section~\ref{matmeth} (since there is no noise with which to compare).
The mixture of populations of unequal strength in the forking fiber
shows a number of characteristics not found in the crossing fiber.  For
example, the forking fiber is clearly diagnosed as a single population
until voxels~(i,e)--(i,g), where there is increasing
heterogeneity in the fiber population.  This is exhibited by increasing
values for the decay ratio $\zeta$-statistic, and the asymmetry
$\kappa$-statistic.  For the crossing fiber, we clearly detect the
multiple-fiber population in voxel~(iii,d) using either the $\tau$ or
$\tilde\tau$ statistics.  The multiple-fiber characteristics in the
forking example are more complex, where the second fiber population is
initially dominated by the first.  If we examine the crossing fiber
more closely, there is little apparent asymmetry and we can compare the
asymmetry statistic, where $\kappa\approx{0}$ versus
$0.15\leq{\kappa}\leq0.45$ for voxels~(i,c)--(i,e).  To distinguish
multiple fibers from uniformity, we observe that $\xi<1$, which is the
expected value under the hypothesis of isotropy.

\section{Simulation study}
\label{sec:simulation}

We illustrate the properties of the proposed $q$-space summary
statistics for the diffusion~PDF on a variety of simulated diffusions
processes.  The following six models attempt to cover common, and not
so common, diffusion processes that include both unimodal and multiple
tensors:
\begin{eqnarray} \label{e:A5}
  \cA_i(\q) &=& \exp(-t \q^T \wt{\bld{D}}_i \q),\qquad  i=1,2,3,\nonumber\\
  \wt{\bld{D}}_1 &=& 68 \tilde\be_1 \tilde\be_1^T + 8\tilde\be_2
  \tilde\be_2^T + 8\tilde\be_3 \tilde\be_3^T,\nonumber\\ [-8pt]\\ [-8pt]
  \wt{\bld{D}}_2 &=& 68 \tilde\be_1 \tilde\be_1^T + 15\tilde\be_2
  \tilde\be_2^T + \tilde\be_3 \tilde\be_3^T,\nonumber\\
  \wt{\bld{D}}_3 &=& 28\tilde\be_1 \tilde\be_1^T + 28\tilde\be_2
  \tilde\be_2^T + 28\tilde\be_3 \tilde\be_3^T,\nonumber
  \end{eqnarray}
  \begin{eqnarray}
  \cA_4(\q) &=& 0.5 \exp(-t \q^T \wt{\bld{D}}_1 \q) + 0.5 \exp(-t \q^T
  \wt{\bld{D}}_4 \q),\nonumber\\ [-8pt]\\ [-8pt]
  \wt{\bld{D}}_4 &=& 68\tilde\be_2 \tilde\be_2^T + 8\tilde\be_1
  \tilde\be_1^T + 8\tilde\be_3 \tilde\be_3^T,\nonumber
  \end{eqnarray}
  \begin{eqnarray}
  \hspace*{15pt}
  \cA_5(\q) &=& \exp(-11t |\q^T\tilde\be_2|^2)\nonumber\\
  &&{}\times\bigl|\exp(-68t |\q^T\tilde\be_1|^2)
  \times [\exp(-0.2t |\q^T\tilde\be_3|^2) +
    \exp(-35t |\q^T\tilde\be_3|^2)]\\
 &&\phantom{{}\times\bigl|}{}+ \frac{4}{\pi}D\bigl(\sqrt{68t} \q^T \tilde\be_1\bigr)
  \bigl[D\bigl(\sqrt{35t} \q^T \tilde\be_3\bigr) -
  D\bigl(\sqrt{0.2t}\q^T \tilde\be_3\bigr)\bigr]\bigr|,\nonumber
  \end{eqnarray}
  \begin{eqnarray}
  \cA_6(\q) &=&0.3\exp(-t \q^T \wt{\bld{D}}_1 \q)+0.7
  \exp(-t \q^T \wt{\bld{D}}_5 \q),\nonumber\\ [-8pt]\\ [-8pt]
  \wt{\bld{D}}_5 &=& 42.5\breve\be_1\breve\be^T_1 + 14\breve\be_2
  \breve\be^T_2 + 20\breve\be_3 \breve\be^T_3,\nonumber
\end{eqnarray}
where $D(x)=\exp(-x^2)\int_0^x\exp(t^2)\,dt$ is the Dawson function
[\cite{abra}].  We define $\tilde\be_j=\R\be_j$, where the matrix $\R$
rotates the axes $(\be_1,\be_2,\be_3)$ to a new coordinate system
$(\tilde\be_1,\tilde\be_2,\tilde\be_3)$.  This extra step is added to
protect against systematic bias in our estimation procedure due to the
diffusion~PDF coinciding with the sampling grid.  In $\cA_6(\q)$ this
rotation is not implemented, but
$(\breve\be_1,\breve\be_2,\breve\be_3)$ has been rotated with respect
to $(\be_1,\be_2,\be_3)$ to produce an asymmetric diffusion in the
multi-tensor model.

These diffusion processes are displayed in both spatial and frequency
domains in Figure~\ref{fig:inrha}, where $\cA_1(\cdot)$ is a prolate
diffusion model (a and g), $\cA_2(\cdot)$
is a scalene diffusion (b and h),
$\cA_4(\cdot)$ is a mixture of two crossing fibers (c~%
and i), $\cA_6(\cdot)$ is the first asymmetric diffusion
(d and j), and $\cA_5(\cdot)$ is the
second asymmetric diffusion (e and k).
For completeness an isotropic diffusion model $\cA_3(\cdot)$ is shown
in Figure~\ref{fig:inrha}f and l.

\begin{table}
  \caption{Hypothesis tests for the six diffusion processes
    $\{\cA_i\}_{i=1}^6$, specified in \protect\eqref{e:A5}, where the number
    of rejected hypothesis are provided as a single number out of 1000
    tests, or as a~fraction if fewer than 1000 tests were performed.
    The nominal size of the tests is $5\%$ for $U$, $V$ and $Q$, while
    the nominal size is $10\%$ for $K$ and $\wt{U}$.  The hypothesis
    tests have been carried out at different SNRs, where the noise
    standard deviation is increasing as you go further down the
    entries $\mbox{SNR}\in\{1/30,1/20,1/10,1/2\}$.  Keys to the
    abbreviations are N-P$/$A (nonpreference versus anisotropy), C$/$E
    (circular versus ellipsoidal), S$/$A (symmetric versus asymmetric),
    I$/$M (isotropic versus multi-modal) and M$/$U (multi-modal versus
    unimodal)}\label{testing}
    \vspace*{-3pt}
  \begin{tabular*}{\textwidth}{@{\extracolsep{\fill}}lccccccc@{}}
    \hline
    $\bolds{H_0/H_1}$ & \textbf{Statistic} & $\bolds{\cA_1}$ & $\bolds{\cA_2}$
    & $\bolds{\cA_3}$ & $\bolds{\cA_4}$ & $\bolds{\cA_5}$ & $\bolds{\cA_6}$\\
    \hline
    \multicolumn{8}{@{}l@{}}{$\mbox{SNR}=1/30$}\\
     N-P$/$A & $U$ & 1000 & 988 & 26 & 492 & 1000 & 802\\
    C$/$E & $V$ & 146$/$1000 & 924$/$988 & 0$/$26 & 382$/$492 & 120$/$1000 & 495$/$802\\
    S$/$A & $K$ & 191$/$1000 & 108$/$988 & 10$/$26 & 258$/$492 & 491$/$1000& 208$/$802 \\
    I$/$M & $Q$ & 0$/$0 & 12$/$12 & 21$/$974 & 420$/$508 & 0$/$0 & 195$/$198\\
    M$/$U & $\wt{U}$ & 991$/$1000 & 136$/$988 & 0$/$26 & 239$/$492 & 996$/$1000 &
    19$/$802\\ [4pt]
    \multicolumn{8}{@{}l@{}}{$\mbox{SNR}=1/20$}\\
     N-P$/$A & $U$ & 1000 & 855 & 26 & 484 & 1000 & 727\\
    C$/$E & $V$ & 153$/$1000 & 794$/$855 & 0$/$26 & 338$/$484 & 148$/$1000 & 267$/$727\\
    S$/$A & $K$ & 148$/$1000 & 38$/$855 & 10$/$26 & 199$/$484 & 331$/$1000 & 136$/$727\\
    I$/$M & $Q$ & 0$/$0 & 0$/$145 & 23$/$974  &259$/$516& 0$/$0 & 201$/$273\\
    M$/$U & $\wt{U}$ & 945$/$1000 & 46$/$805 & 0$/$23 & 151$/$484 & 942$/$1000 &
    8$/$727\\ [4pt]
     \multicolumn{8}{@{}l@{}}{$\mbox{SNR}=1/10$}\\
      N-P$/$A & $U$ & 1000 & 239 & 34 & 441 & 998 & 457\\
    C$/$E & $V$ & 225$/$1000  & 214$/$239 & 0$/$34 & 192$/$441 & 194$/$998 & 58$/$457\\
    S$/$A & $K$ & 89$/$1000   & 1$/$239 & 10$/$34 & 99$/$441 & 139$/$998 & 74$/$457\\
    I$/$M & $Q$ & 0$/$0 & 449$/$761 & 20$/$966  & 20$/$539&1$/$2  &18$/$543 \\
    M$/$U & $\wt{U}$ & 239$/$1000 & 7$/$239 & 0$/$34 & 27$/$441 & 174$/$998  &
    4$/$457\\ [4pt]
     \multicolumn{8}{@{}l@{}}{$\mbox{SNR}=1/2$}\\
    N-P$/$A & $U$ & 45 & 21 & 25 & 31 & 37 & 24\\
    C$/$E & $V$ & 2$/$45 & 2$/$21 & 5$/$25 & 6$/$31 & 4$/$37 & 4$/$24\\
    S$/$A & $K$ & 5$/$45 & 4$/$21 & 2$/$25 & 1$/$31 & 4$/$37 & 3$/$24\\
    I$/$M & $Q$ & 1$/$955 & 3$/$979 & 2$/$975 &2$/$969& 4$/$963 &5$/$976 \\
    M$/$U & $\wt{U}$ & 2$/$45 & 0$/$21 & 1$/$25 &0$/$31  & 5$/$37  & 1$/$24\\
    \hline
  \end{tabular*}
  \vspace*{-3pt}
  \end{table}

We have chosen to define $\wt{\bld{D}}_i=4\times10^{10}\bld{D}_i$,
$i=1,\dots,4$, and normalized\break $\|\q\|=1$.  With $t=0.04$ this
corresponds to $b=4t\times10^{10}=1600\ \mathrm{s}/\mathrm{mm}^2$ and the trace of
the first three nonnormalized matrices $\bld{D}_i$ as
$2.1\times10^{-9}\ \mathrm{m}^2/\mathrm{s}$ [\cite{Alexander2005}].  The
function $\cA_5(\q)$ is obtained from the magnitude of the FT of an
asymmetrically decaying diffusion process in space.  We illustrate a
range of behavior for the scalar statistics defined in $q$-space using
these test functions, providing only a subset in order to compare and
contrast their performance.  We simulate 1000 realizations for each
test function and add Gaussian noise with standard deviation of
$\cA(0)/2$, $\cA(0)/10$, $\cA(0)/20$ and $\cA(0)/30$ to both the real
and imaginary channels using a 60-direction HARDI sampling scheme.

Results, provided in Table~\ref{testing}, are consistent with varying
degrees of the SNR.  The prolate diffusion $\cA_1$ is clearly
detectable, down to an $\mbox{SNR}=1/10$, despite using nonparametric
methods via the $U$-statistic.  Detecting the scalene diffusion depends
on the SNR, while the isotropic diffusion is clearly distinguishable
from its alternatives under the full range of SNR using the
$U$-statistics.  The multi-tensor diffusion $\cA_4$ is difficult to
classify using the $U$-statistics and its correct classification
depends on how well the location of the dominant peak is estimated.  If
the dominant peak is well determined, then the $U$-statistic clearly
recognizes the density as anisotropic, if not, the $q$-space
measurements are characterized as non-Gaussian instead of multi-modal.
If one was only concerned with empirically separating prolate
diffusion~PDFs from multi-modal diffusion~PDFs, rather than performing
a~hypothesis test, then this would be relatively straightforward, for
example, retaining 95\% of the unimodal Gaussian with the
$\mbox{SNR}=1/20$ leads to rejecting all but 11\% of the multi-tensor
realizations (see the $\tilde{U}$-statistic).  Since we are interested
in detecting ellipsoidal decay around a single direction, the variation
over the dominant great circle will be large for anisotropic voxels
with ellipsoidal decay\vspace*{1pt} as well as for multi-modal diffusion~PDFs.  At
an $\mbox{SNR}=1/20$ the $\wt{U}$-statistic provides complimentary
information by strongly separating the prolate (94\% rejected) from the
multi-tensor model (15.1\% rejected, near the nominal value of 10\%),
but fails to distinguish between the scalene and the multi-tensor
models (Table~\ref{testing}).  The highly scalene diffusion is mistaken
(not surprisingly) for a~multi-modal diffusion and such structure may
be approximated using two tensors, especially when sparsely sampled on
the sphere.

The two distributions with constant behavior on the dominant great
circle are not diagnosed with asymmetric decay, while the null
hypothesis is rejected for $\cA_2$ in a substantial number of cases in
Table~\ref{testing}.  The misdiagnosed multi-modal diffusion~PDF
$\cA_4$ also has the null hypothesis of multimodality rejected for
a~substantial number of cases.  This is to be expected since the observed
diffusion will experience considerable variation over the dominant
great circle, consistent with observing a diffusion process with a
single dominant direction and ellipsoidal decay.

We fail to reject the null hypothesis of symmetry for the two diffusion
processes that are symmetric ($\cA_1$ and $\cA_2$) in most cases, while
we reject a larger proportion for $\cA_5$.  There is unfortunately a
lack of power in this test which is due to sampling 60~directions,
limiting the performance of the test statistic.  For $\cA_6$ and
keeping the $\mbox{SNR}=1/20$, we reject the null hypothesis 38.9\%
with 60~directions.  For $\cA_5$ we reject the null 51.6\% of the time
using 245~directions at $\mbox{SNR}=1/20$---a clear increase from
35.2\% with 60~directions.  Increasing the SNR also increases our power
to detect such asymmetry, as shown in Table~\ref{testing}.  The power
of the test improves as the number of directions increase or the amount
of asymmetry (better characterized with better spherical sampling)
increases.  For all of the structural tests performed here there is a
direct similarity in effect of increasing the number of grid points to
improve the size of the mean under the alternative hypothesis or
directly decreasing the variance. This is because as the mean of the
alternative hypothesis increases with improved sampling of $q$-space,
this has the same effect as increasing the SNR, as the test statistics
are (approximately) functions of their ratio.  This direct
exchangeability of sampling in frequency versus SNR holds until the
distributional approximations break down because of poor resolution in
$q$-space or a diminished signal-to-noise ratio.

\begin{figure}

\includegraphics{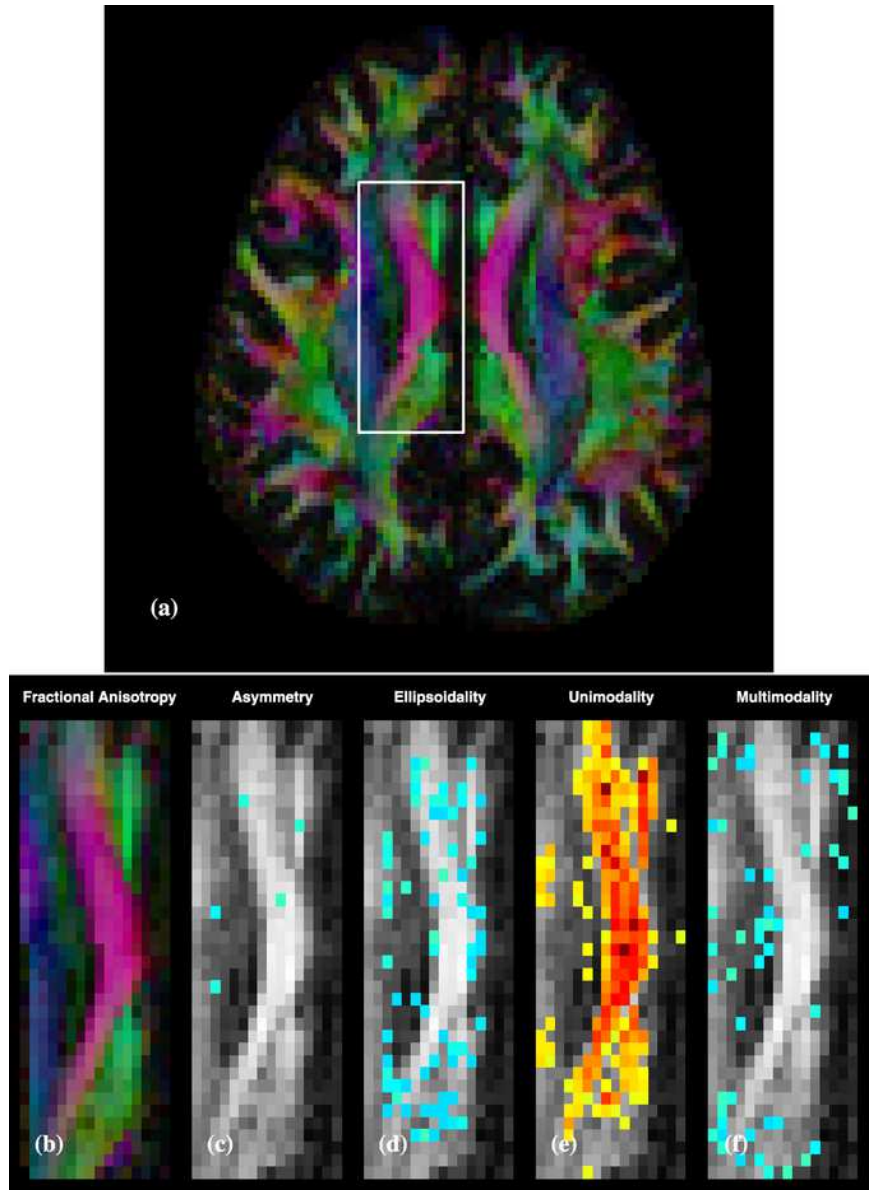}

  \caption{Axial slice from clinical HARDI acquisition.  Color-coded
    fractional anisotropy (FA) for the whole slice is displayed in
    \textup{(a)} along with the boundaries for the ROI (Regions of
    interest).  For the zoomed-in ROI: color-coded FA \textup{(b)},
    anisotropy $p$-values \textup{(c)}, ellipsoidality $p$-values
      \textup{(d)}, unimodality test statistic \textup{(e)} and multimodality
    $p$-values \textup{(f)}.}\label{figure-slice1}
\end{figure}

\section{Analysis of clinical data}

HARDI data were acquired from one normal subject (30~year old,
male~Caucasian) in a Siemens TIM Trio 3.0~Tesla scanner using a
32-channel head coil.  Measurement of 64 gradient directions
($b=1600\ \mathrm{s}/\mathrm{mm}^2$) and one T2 image ($b=0$) were obtained using a
twice-refocused diffusion preparation.  The slice prescription was
64~slices acquired in the AC--PC plane, $\mbox{TE}=95~\mbox{ms}$,
$\mbox{FoV}=240\times240~\mbox{mm}$, $\mbox{base
  resolution}=128\times128$, slice thickness of $1.9~\mbox{mm}$ and
cardiac gating was applied.

Regions of interest (ROIs) from two slices of the clinical data are
provided to illustrate the statistical summaries developed in this
paper.  Slice~1 contains an ROI that is dominated by single-fiber
voxels containing structures such as the corpus callosum and cingulum.
Figure~\ref{figure-slice1}a shows the voxels using the common
color-coding convention [i.e., RGB for the $(x,y,z)$ coordinates]
weighted by the estimated fractional anisotropy (FA) at each voxel. The
FA for the ROI is reproduced in Figure~\ref{figure-slice1}b along with
the $p$-values for the anisotropy and ellipsoidality statistics in
Figure~\ref{figure-slice1}c and d, respectively.
We select a very liberal threshold ($p=0.15$) for the purpose of
exploratory data analysis, not confirmatory data analysis. We observe
very few voxels that indicate asymmetry at specific voxels, while the
ellipsoidality $p$-values indicate quite a few voxels that exhibit
prolate diffusion.  These voxels are located at the borders of strongly
directional structures such as the corpus callosum and cingulum, and
reaffirm the results obtained in the simulation studies.  Additional
information about the structure is obtained by plotting the test
statistic for unimodality and the $p$-values from the multi-modality
test statistic in Figure~\ref{figure-slice1}e
and f, respectively.  The corpus callosum, and to a
lesser extent the cingulum, produce large values in the unimodality
test statistic as to be expected from those structures.  Multimodality
is detected in voxels with reduced FA and/or on the edges of prominent
white-matter structures.  The pattern of multi-modal voxels identified
in Figure~\ref{figure-slice1}f in general do not appear to overlap with
those voxels that were identified using the ellipsoidality statistic,
providing evidence that this methodology is detecting distinct features
in the white-matter microstructure.

\begin{figure}

\includegraphics{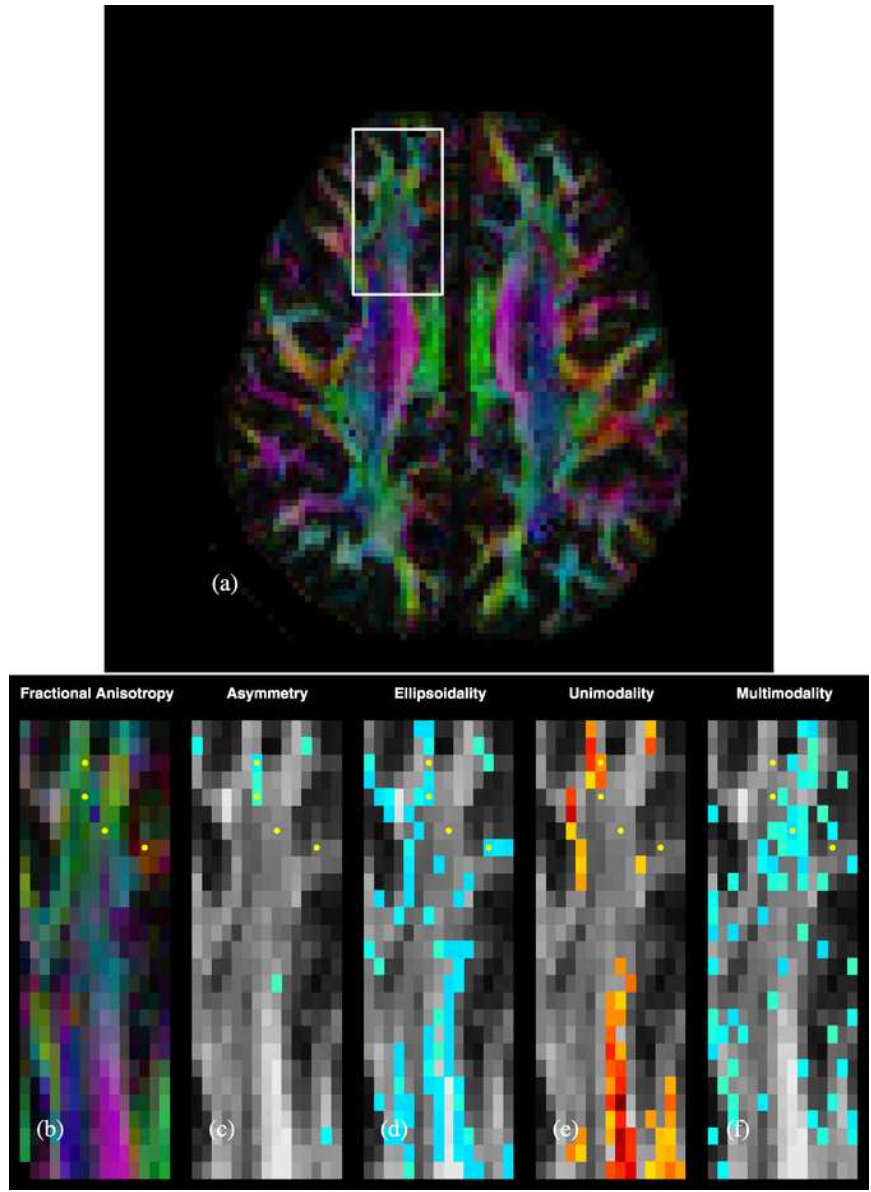}

  \caption{Axial slice from clinical HARDI acquisition.  Color-coded
    fractional anisotropy (FA) for the whole slice is displayed in
    \textup{(a)} along with the boundaries for the ROI.  For the
    zoomed-in ROI: color-coded FA \textup{(b)}, anisotropy $p$-values
    \textup{(c)}, ellipsoidality $p$-values \textup{(d)}, unimodality test
    statistic \textup{(e)} and multimodality $p$-values \textup{(f)}.} \label{figure-slice2}
\end{figure}

The ROI selected in slice~2 captures more complicated interactions
between white-matter structures such as the corticopontine tract,
anterior thalamic radiation and corpus callosum
(Figure~\ref{figure-slice2}a).  The FA for the ROI is reproduced in
Figure~\ref{figure-slice2}b along with the $p$-values for the
anisotropy and ellipsoidality statistics in
Figure~\ref{figure-slice2}c and d, respectively.
Asymmetry is difficult to detect in these data, but ellipsoidality is
quite apparent along the boundaries of the corpus callosum and around
the projections into gray matter.  The test statistic for unimodality
in Figure~\ref{figure-slice2}e complement the ellipsoidality results
quite well, picking out dominant prolate diffusion (e.g., the voxels
dominated by the corpus callosum and to a lesser degree the cingulum)
around which the ellipsoidality measure is finding more complex voxels.
Finally, the test statistic for multimodality in
Figure~\ref{figure-slice2}f clearly identifies voxels where the three
dominant white-matter structures in this ROI converge, and all other
statistics fail to detect any specific structure.  The statistical
summaries developed here provide complementary information about
white-matter microstructure in clinically acquired data.

\begin{figure}

\includegraphics{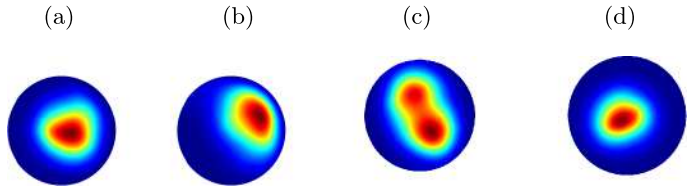}

  \caption{The raw Funk Radon Transform (FRT) from a collection of
    voxels indicated by yellow dots in Figure~\protect\ref{figure-slice2}.
    These are plotted in order of decreasing $x_2$-coordinate (or
    going from the top of the image to the bottom).  Subplots \textup{(a)} and
    \textup{(b)} both reject the null hypothesis of no asymmetry, with \textup{(a)} not
    rejecting prolate diffusion in favor of scalene diffusion.
    Subplot \textup{(c)} rejects unimodality in favor of multimodality and also
    rejects isotropy in favor of multimodality.  Subplot \textup{(d)} is
    unimodal.  The raw FRTs are consistent with these diagnoses.}\label{fig:indvoxel}
\end{figure}

We focus on a few specific voxels in Figure~\ref{figure-slice2} using
the Funk--Radon Transform (FRT) without smoothing.  As recommended by
\cite{Tuch}, we have taken the standardized raw FRT to the power five
to emphasize structure in the display.  Figure~\ref{fig:indvoxel}a
and b show the two most anterior voxels that are
plotted in Figure \ref{figure-slice2} (indicated by yellow dots).  This tract
appears to be ``bending'' as we move from anterior to posterior,
indicated by the shift in direction of the dominant direction seen in
the FRTs.  The statistics quantify this behavior; the $p$-values for
asymmetry are 0.14 and 0.02 respectively (indicating that the
posterior-most voxel is bending more).  The unimodality of the anterior
voxel is seen from the large unimodality statistic in
Figure~\ref{figure-slice2}e.  We then look at a voxel in a more
heterogeneous area, where the major fiber tracts appear to merge: the
statistics here indicate multi-modality dominates as is seen in
Figure \ref{fig:indvoxel}c and backed up by Figure~\ref{figure-slice2}c--f.  We observe the most central voxel has
summary statistics that are ellipsoidal but not asymmetric, clearly
observed in Figure~\ref{fig:indvoxel}d.  The clinical data have
provided evidence at a~voxel level, backed up by statistical hypothesis
testing and observed in the FRT visualizations, that interesting
white-matter microstructure may be detected and characterized using the
methodology proposed here.

\section{Discussion}

We have introduced a new set of tools for characterizing orientational
structure from HARDI measurements directly in $q$-space.  This
methodology is unique when compared with existing methods that rely on
reconstructing the spatial information from $q$-space by different
methods of marginalizing the spatial distribution, that is, from
calculating a spatial ODF.  An ODF has a different meaning if
calculated directly from a Gaussian model, from the nonparametric FRT
(average orientational distribution over all radii without using the
correct volume increment for a marginal PDF) or using PAS-MRI
(orientational distribution associated with a single spatial radius or
scale).  In general, the magnitude associated with an ODF is not
comparable between methods, neither is the distribution of noise
artifacts.  Our methodology is technically linked to the FRT, but
unlike the FRT we are not constrained to scalar measures calculated
from averages on great circles in $q$-space, and our methods do not
depend on appropriate marginalization to produce summaries.  The
interpretation of our statistical summaries is straightforward, but we
note that in improvements in data acquisition, such as increased
sampling of directions.

Most established methods for characterizing features in white-matter
microstructure have focused on the problem of determining the number
and orientation of peaks in the diffusion~PDF.  None of the
``magnitude'' information of these solutions is comparable or indeed
interpretable apart from DTI-based models.  \cite{Savadjiev2006} have
already commented on the unsuitability of such magnitudes as
quantitative measures.  The problem with this fact, and the nonlinear
transformation often employed for representing $q$-ball estimates, is
that the coherent treatment of noise artifacts becomes much more
difficult.  The advantage of our theoretical framework, developed for
summary statistics, is that we may perform hypothesis tests using
critical values that are not functions of unknown parameters.  We
stress that simulation studies for features of diffusion~PDFs are in
general misleading unless the proposed summaries are true statistics,
that is, their distributions under null hypotheses are parameter
independent.  For example, critical values determined from Monte Carlo
studies for a given diffusion~PDF will not (in general) be applicable
to other diffusion processes than the simulated process since these
critical values are parameter dependent.  This can be compared to
calculating a simple mean rather than a $t$-statistic.  If we try to
elicit the distribution of the sample mean using simulations at fixed
variances, then these critical values are only useful for variables
with the same variance.

Various nonparametric procedures have been proposed to summarize\break HARDI
data using more than its estimated orientation, for example, by
investigating the model order of spherical harmonic decomposition
[\cite{fracharacterization}; \cite{alebararrdetection}; \cite{Descoteaux2006}].
\cite{chen2005} modeled the ADC using a product of a truncated
spherical harmonics series.  In general, an infinite order of spherical
harmonic terms must be taken to approximate an arbitrary Gaussian
mixture, but they argued that a crossing fiber should be sufficiently
reproduced by such a truncated representation and expressed its
complexity using the normalized terms in the spherical expansion.
Other representations include expressing the ADC in terms of
higher-order tensors and spherical harmonics [\cite{Descoteaux2006}],
or just via a spherical harmonic representation
[\cite{fracharacterization}].
Second-order terms in a spherical harmonic decomposition contribute to
describing a single-tensor fiber, but more complicated structure must
be described in terms of corresponding spatial properties of the PDF
directly, rather than the fourth- and higher-order terms which give too
much freedom in structure to be a precise tool for the description of
fine spatial features.  Other measures of the entropy of the
diffusion~PDF have been proposed by \cite{rao2004}.

Rather than solely focusing only on the number of peaks in the
diffusion~PDF, we have characterized white-matter microstructure
through the diffusion~PDF directly in $q$-space without parametric
assumptions or imposing smoothness constraints, as we use a variable
bandwidth estimator rather than employing a fixed bandwidth smoother
[\cite{OlhedeWhitcher}].  The tissue microstructure is identified as
variation in summary statistics that deviate from a simple, symmetric
model for the diffusion~PDF and is characterized in behavior relative
to the identified dominant great circle in $q$-space.  Ellipsoidal
diffusion~PDFs \eqref{ellipsoid} are simple in structure and imply the
existence of a dominant great circle.  The deformed ellipsoid class is
less stringent in structure, and permits asymmetric decay in minor
axes---for example, \eqref{deformedell}---while still conforming to the
existence of a dominant great circle.  We describe the precursor to
forking structures by either a deformed ellipse or a mixture model, to
capture further asymmetric structure.  We differentiate between
different white-matter microstructure by examining variation over that
great circle, or variation perpendicular to the great circle. Allowing
for a greater variety of structure in a unidirectional diffusion~PDF
implies that the power to detect multi-modal diffusion is necessarily
reduced compared to using a parametric multi-model model, if the
proposed parametric model is correct.  We characterized single peak
densities by additional summaries, such as the anisotropy statistic,
the decay ratio statistic and the asymmetry statistic.  The synthetic
forking fiber in Figure~\ref{fig:evolution} shows an evolution of such
measures as we go between a single fiber and a forking fiber.  The
synthetic crossing fiber in Figure~\ref{fig:evolution} does not exhibit
the same asymmetries.

If one enforces a strict Gaussian (single diffusion tensor) model, then
all variation away from symmetry around the dominant direction will be
interpreted as evidence for a multi-modal diffusion
[\cite{paralepico}; \cite{hoswilansinference}; \cite{Behrens2007}].  Modeling
using non-Gaussian PDFs allows us to fit asymmetric structure, rather
than just the model indicating a lack of fit of a single peak. However,
using such models leads to a loss of power if a Gaussian mixture model
is appropriate.  Caution should be exercised in order to protect
against over-interpreting fitted models.  With a model that only
includes a family of mixtures of Gaussian diffusion processes, one is
constrained to estimate a Gaussian mixture, however, for a small number
of sampled directions there will inevitably be issues with
identifiability.  The same realizations may in some cases equivalently
be derived from a unimodal diffusion~PDF with asymmetric structure or a
Gaussian mixture model.  If one chooses to select one model rather than
the other (i.e., choose an asymmetric and scalene PDF or
multiple-tensor), then this decision is based more on the underlying
assumptions of the model rather than on the evidence directly provided
by the observed data.  A large (possibly infinite) collection of
Gaussian diffusion processes may be used to approximate an observed set
of measurements to an arbitrary accuracy, but one has to consider the
possibility that the information being fitted is noise instead of
signal.  We believe the rule of parsimony should be exercised at all
times, and that summaries of orientational structure can be estimated
and interpreted in $q$-space rather than using (potentially)
over-parameterized models.

One potential extension to the methods proposed here would be to
acquire multiple shells of a fixed radius in $q$-space instead of
typical HARDI sampling [\cite{Wu2007}; \cite{Wisco07}], that is,
multiple-wavevector or hybrid imaging.  In this case the test
statistics are calculated for each shell, and then averaged across the
different shells.  The dominant orientation would be estimated by a
weighted averaging of the estimated dominant orientations for each
shell, since its distribution depends on the SNR that is
shell-dependent.  Another possible acquisition method is diffusion
spectrum imaging (DSI), corresponding to a Cartesian sampling of the
characteristic function [\cite{Wedeen05}].  It is more difficult to
achieve the same directional resolution in DSI versus
multiple-wavevector imaging, and so with realistic sampling times it
may not be feasible to perform the same analysis as outlined in this
paper.  However, other nonparametric summaries could be defined
directly in $q$-space to characterize the spatial properties.

One potential application of these $q$-space summaries would be to
improve fiber-tracking algorithms, similar to the use of the Hessian of
a local peak to improve probabilistic tractography models
[\cite{Seunarine}].  These summaries would be used in addition to
directions, to allow more careful tracking through forking and fanning
structures (Figure~\ref{figure-slice2}), and distinguish local
structure more consistently with crossing from such features using both
the asymmetry and ellipsoidality measures.

\section*{Acknowledgment}
The authors thank
an anonymous reviewers for a careful reviewing
of the paper.

\begin{supplement}[id=suppA]
  \sname{Supplement}
  \stitle{Distributions for test statistics\\}
  \slink[doi, text={10.1214/10-AOAS441 SUPP}]{10.1214/10-AOAS441SUPP}
  \slink[url]{http://lib.stat.cmu.edu/aoas/441/supplement.pdf}
  \sdatatype{.pdf}
  \sdescription{The supplementary material would be provided at this location.}
\end{supplement}

\printaddresses

\end{document}